\documentclass[epsfig,useAMS,usenatbib]{mn2e}
\usepackage[dvips]{graphicx}
\input{epsf}
\title[Anomalous CMB variance]{Anomalous variance in the WMAP data and Galactic Foreground residuals}
\author[M. Cruz et al.]{M. Cruz,$^{1}$\thanks{E-mail:
marcos.cruz@unican.es} 
P. Vielva,$^{2}$  %\footnotemark[1]
E. Mart\'{\i}nez-Gonz\'alez,$^2$
R. B. Barreiro$^2$
\\
$^{1}$Depto. de Matem\'aticas, Estad\'{\i}stica y Computaci\'on, Universidad de Cantabria, Avda. los Castros, s/n,  39005-Santander, Spain\\
$^{2}$IFCA, CSIC-Univ. de Cantabria, Avda. los Castros, s/n, 39005-Santander,Spain}
\begin{document}

\date{Accepted Received; in original form}

\pagerange{\pageref{firstpage}--\pageref{lastpage}} \pubyear{2010}

\maketitle

\label{firstpage}

\begin{abstract}
A previous work \citep{mon08} estimated the CMB variance from the three-year WMAP data,
finding a lower value than expected from Gaussian simulations using the WMAP best-fit cosmological model. 
We repeat the analysis on the five-year WMAP data using a new estimator with lower bias and variance.
Our results confirm this anomaly at higher significance, namely with a $p$-value of $0.31\%$.
We perform the analysis using different exclusion masks, showing that a particular region of the sky near the Galactic plane shows a higher variance than $95.58\%$ of the simulations whereas the rest of the sky has a lower variance than $99.96\%$ of the simulations. The relative difference in variance between both regions is bigger than in $99.64\%$ of the simulations. This anisotropic distribution of power seems to be causing the anomaly since the model assumes isotropy. Furthermore, this region has a clear frequency dependence between 41GHz and 61GHz or 94GHz suggesting that Galactic foreground residuals could be responsible for the anomaly.
Moreover, removing the quadrupole and the octopole from data and simulations the anomaly disappears.
The variance anomaly and the previously reported quadrupole and octopole alignment seem therefore to be related and could have a common origin. We discuss different possible causes and Galactic foreground residuals seem to be the most likely one.
These residuals would affect the estimation of the angular power spectrum from the WMAP data, which is used to generate Gaussian simulations, giving rise to an inconsistency between the estimated and expected CMB variance. If the presence of residuals is confirmed, the estimation of the cosmological parameters could be affected.
\end{abstract}


\section{Introduction}

The Cosmic Microwave Background (CMB) is the most ancient image of the Universe. This image can reveal valuable information on the origin and evolution of the Universe. Many observations in different cosmological fields tend to support the \emph{concordance}, standard inflationary model. This model predicts the temperature anisotropies of the CMB to represent a Gaussian and isotropic random field on the sphere, whereas non-standard models, such as topological defect models, predict departures from isotropy and/or Gaussianity.

The Wilkinson Microwave Anisotropy Probe (WMAP) satellite, provided the first all-sky, high resolution map 
\citep{ben03} of the CMB anisotropies. 
In the past years, many authors claimed to have found anomalies in the WMAP data. Some of them could be affected by a posteriori statistics but the list of anomalies is too long to be ignored without worry. Some of these anomalies are,
asymmetry between ecliptic hemispheres \citep{eri04a,han04a,eri04b,han04b,eri05,don05,lan05, ber07, eri07, ber09, hof09, pie10, vie10, cay10};
anomalous quadrupole-octupole alignment 
\citep{ bie04, cop04, oli04, sch04, bie05,abr06,cop06,cop07,gru09,fro10}; unexpected alignment of CMB structures towards the Ecliptic poles \citep{wia06, vie07}; a prominent cold spot in the southern Galactic hemisphere \citep{vie04, muk04, cru05, cru06, cru07a, cru07b, cru08, mce05, cay05, rae07, wia08, gur08,pie08,gur09,ros09}.

These anomalies arise after using very sophisticated statistical methods. However, a very simple analysis already reveals some anomalies. \citet{mon08} estimated the CMB variance of the 3--year WMAP data and compared it to the values obtained from Gaussian simulations. The $p$-value of the variance was $1.2\%$. They confirmed this result by comparing with an analytic formula derived in \citet{cay91}. An analysis of the two ecliptic hemispheres revealed that the anomaly was mainly due to the low variance of the Northern ecliptic hemisphere. Similar results were found in \citet{lar04} and \citet{aya09}, analysing spot abundances. 

Here we propose a new method to estimate the CMB variance from data with anisotropic noise. In section 2 we review the WMAP data and exclusion masks we use. We compare the new method to the one used in \citet{mon08} in section 3 and  apply it to the 5--year WMAP data in section 4. Finally we present the conclusions of this paper in section 5.

\section{WMAP data and exclusion masks}

The WMAP team \citep{hin09} provides foreground-reduced maps
for the Q (41 GHz), V (61 GHz), and W (94 GHz) bands.
We use the WMAP foreground cleaned VW combined map:

\begin{equation}
\label{eq:combination}
T_c({\mathbf{x}}) = \sum_{i = 1}^{6} 
{T_i}({\mathbf{x}})~{w_i}({\mathbf{x}}),
\end{equation}
where $\mathbf{x}=(\theta,\phi)$ gives the position in the sky and the index $i$
stands for the 6 differencing assemblies of the V and W bands, namely V1, V2 for the V band and W1, W2, W3 and W4 for the W band.
These are the data recommended for cosmological analyses, since their Galactic foreground contamination should be low.
The noise weight ${w_i}({\mathbf{x}})$ is defined as:
\begin{eqnarray}
\label{eq:noise}
w_i({\mathbf{x}}) = \frac{\bar{w}_i({\mathbf{x}})}{\sum_{i =
1}^{6}{\bar{w}_i}({\mathbf{x}})},~~ &
\bar{w}_i({\mathbf{x}}) = \frac{{N_i}({\mathbf{x}})}{{{\sigma_0}_i}^{2}},
\end{eqnarray}
where ${{\sigma_0}_i}$ is the noise dispersion per observation for
each differencing assembly and ${N_i}({\mathbf{x}})$ is the number of observations made by the
receiver $i$ at the position in the sky ${\mathbf{x}}$.

Although the data are provided at a
HEALPix\footnote{http://healpix.jpl.nasa.gov} \citep{gor05} 
resolution of $n_{side}=512$, we degrade the data to $n_{side}=256$ since the
smallest scales are dominated by noise. In addition, in order to avoid
the strong contamination present at the Galactic plane and the
emission coming from bright extragalactic point sources, we will apply different exclusion masks.

\begin{figure}
  \begin{center}
    \includegraphics[width=4cm]{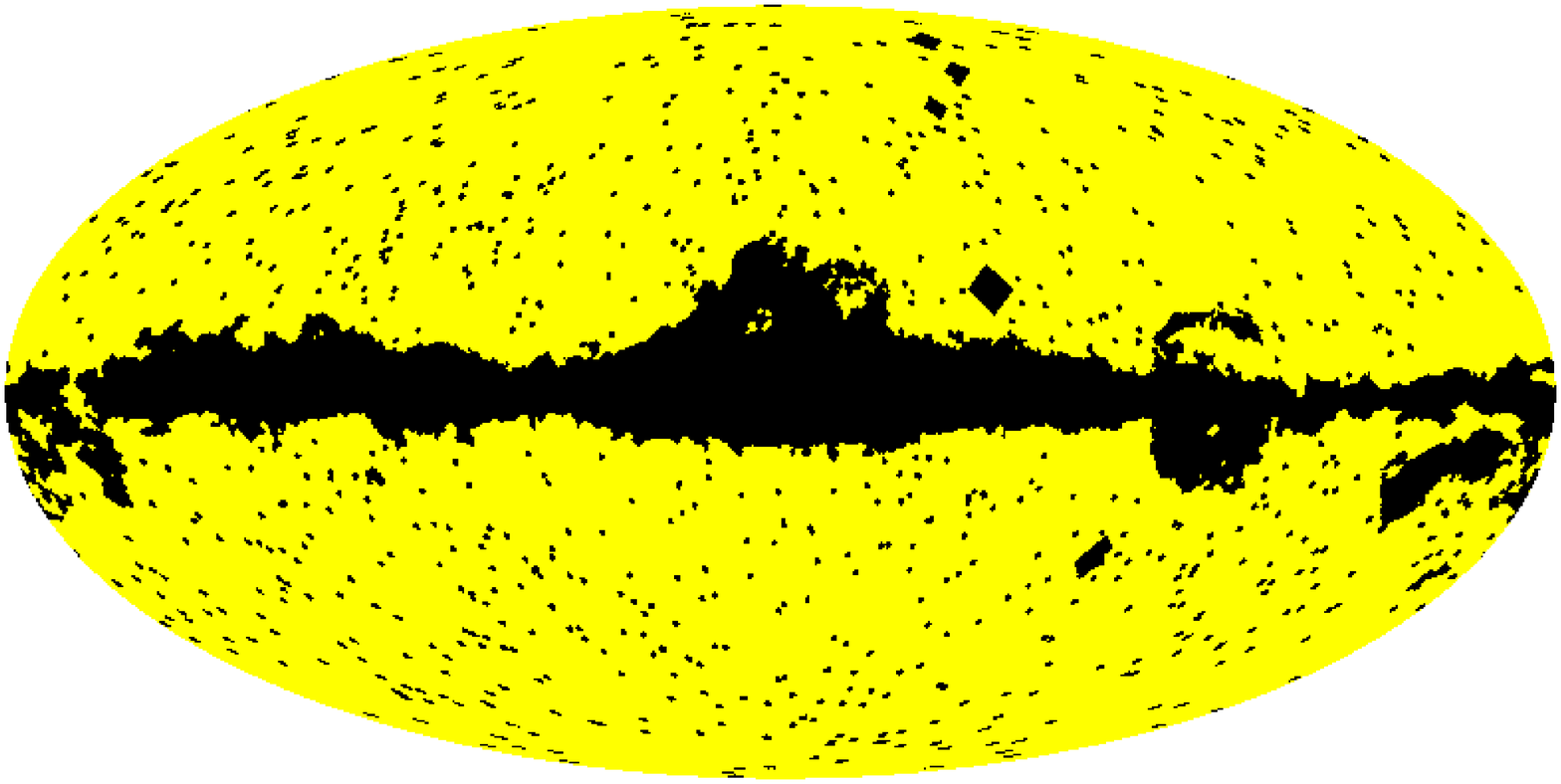}
    \includegraphics[width=4cm]{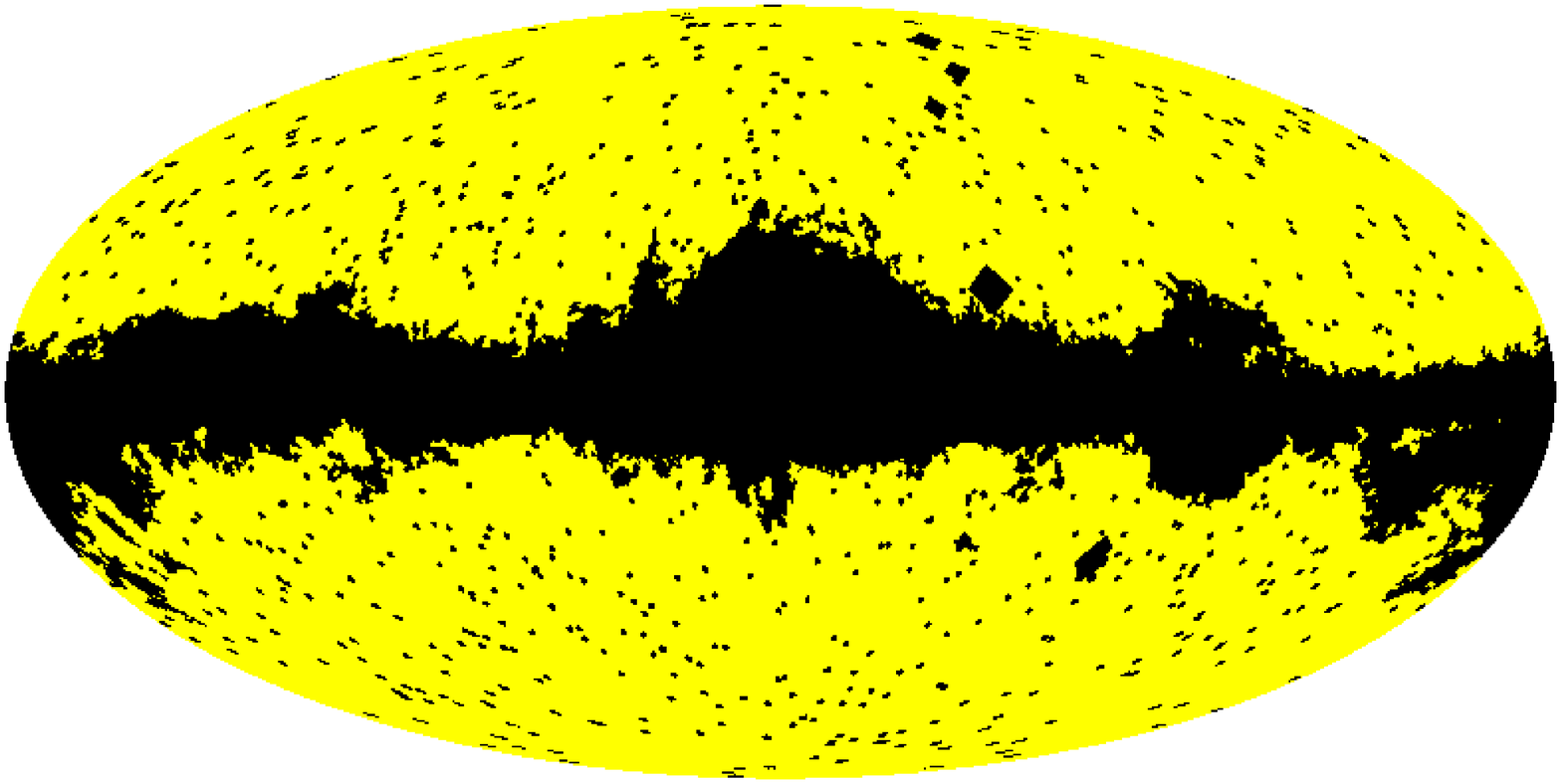}
    \includegraphics[width=4cm]{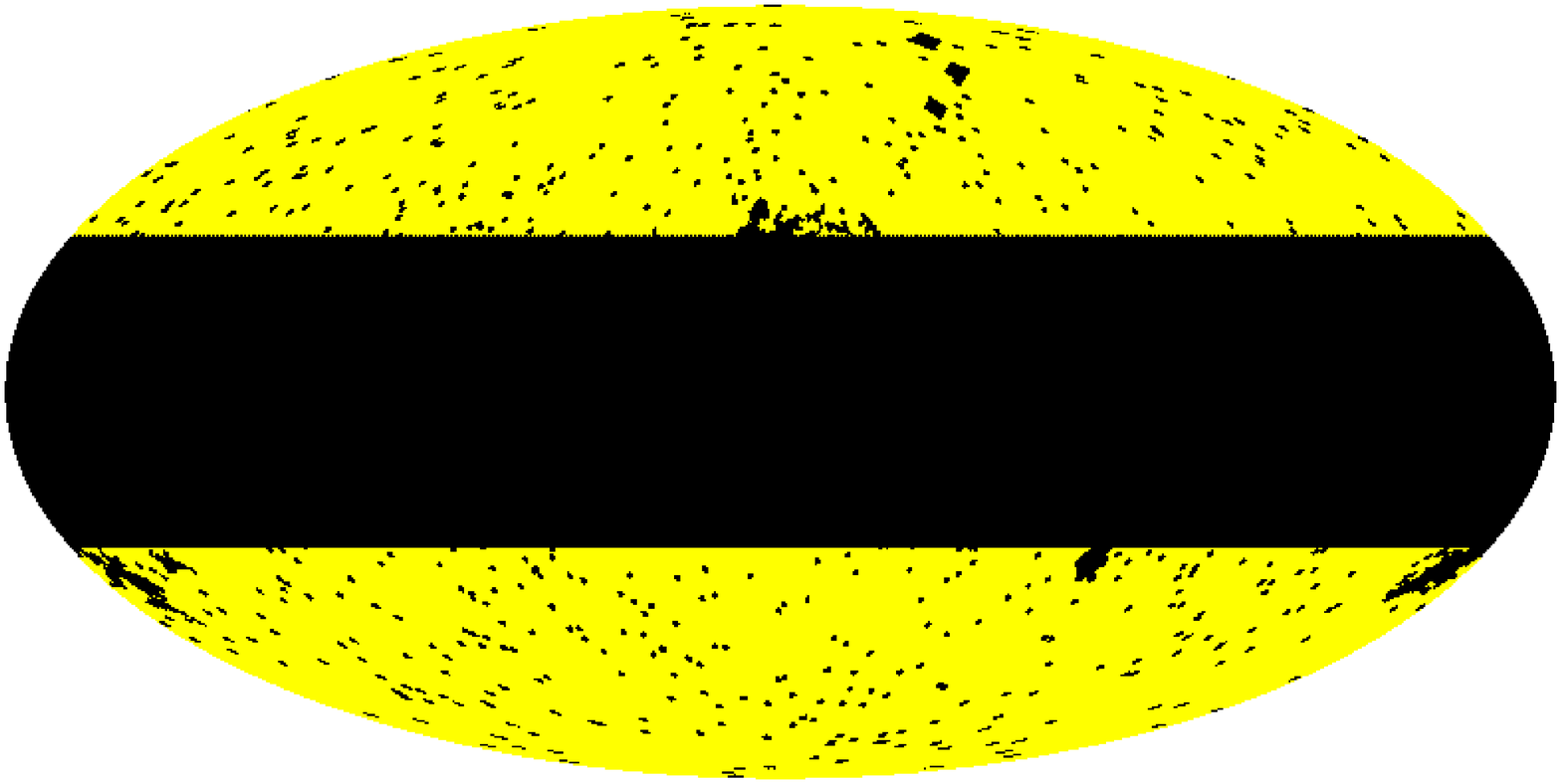}
    \includegraphics[width=4cm]{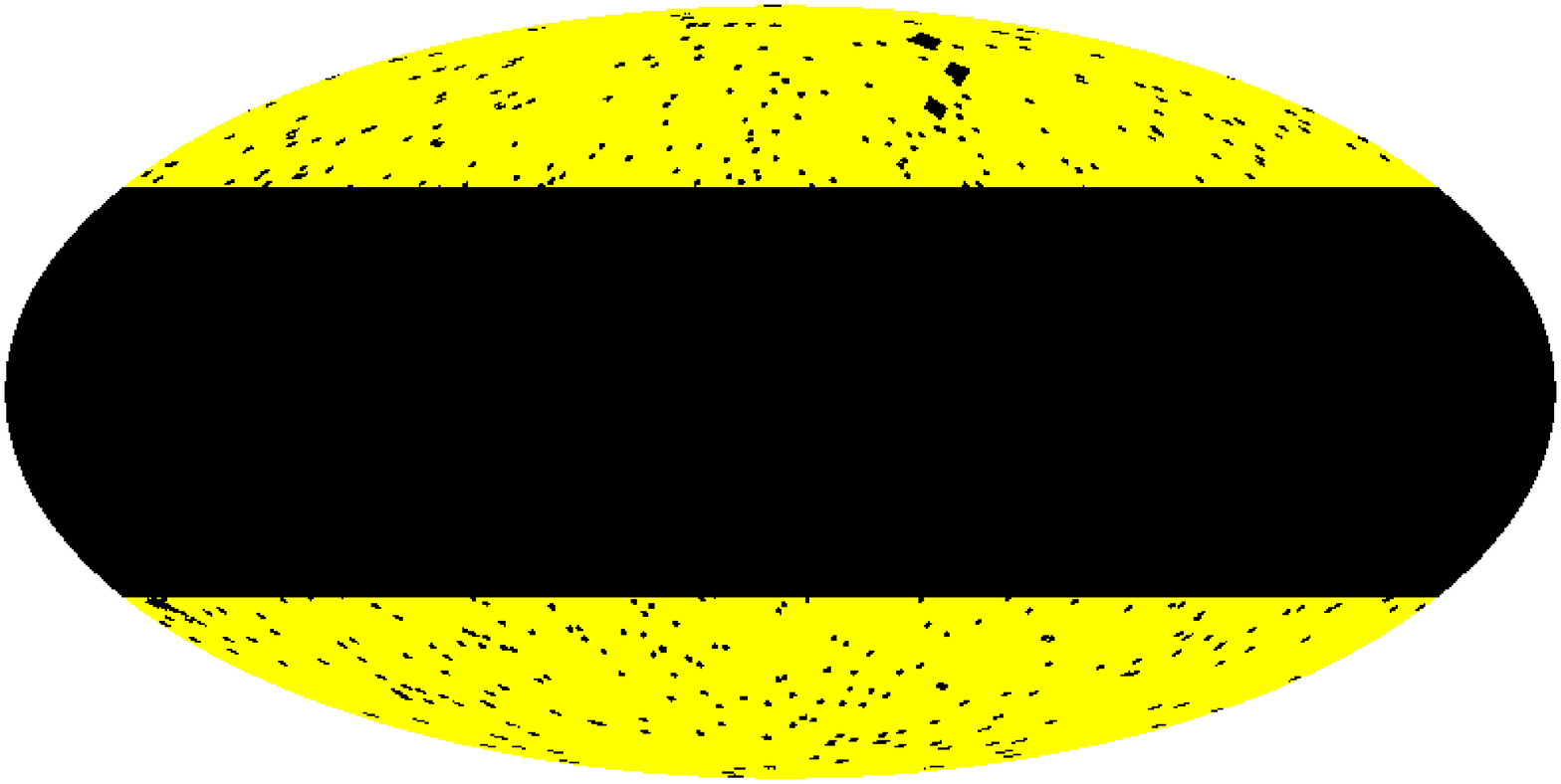}
  \end{center}
    \caption{Mollweide projections of four exclusion masks used in the analysis. From top to bottom and left to right we have the kq85, kq75, gc30 and gc40 masks. The excluded regions are plotted in black.}
   \label{fig1}
\end{figure}

The WMAP team provides the kq85 and kq75 masks which leave approximately $85\%$ and $75\%$ of the sky unmasked.
We construct two more masks by applying an additional $\pm 30$ and $\pm 40$ degrees Galactic latitude cut to the kq75 mask. 
Hereafter we will call these masks gc30 and gc40 respectively.
These four masks are shown in Figure \ref{fig1}

In our Galactic foreground analysis we build the gp33 and gp10 masks which admit only some pixels near the Galactic plane. The description of these masks is given in section 4.

The simulations are generated from the best fit angular power spectrum estimated from the data by the WMAP team. 
Note that the kq85 mask is used in the angular power spectrum estimation \citep{nol09}.
Isotropy and Gaussianity is assumed and the beam and noise properties are taken into account.

Note that the kp0 and kp2 masks given in the 3--year WMAP data release used in \citet{mon08}, are analogous to the kq75 and kq85 masks but slightly different.

\section{CMB variance estimation}

Assuming that the foreground contamination has been totally removed, 
our data, $T(\mathbf{x})$, consist of two components, namely 
CMB, $T_{\mathrm{CMB}}(\mathbf{x})$, and instrumental noise, $T_{\mathrm{noise}}(\mathbf{x})$.

\begin{equation}
T\left(\mathbf{x} \right) = T_{\mathrm{CMB}}\left(\mathbf{x} \right)+T_{\mathrm{noise}}\left(\mathbf{x} \right).
\label{eq:t}
\end{equation}

The standard cosmological model predicts $T_{\mathrm{CMB}}(\mathbf{x})$ to represent an isotropic Gaussian random field. Hence,  the CMB variance, $\sigma_0^2$, 
is constant for any position of the sky, $\mathbf{x}$, according to that model.
Therefore

\begin{equation}
T_{\mathrm{CMB}}\left(\mathbf{x} \right) \sim N\left(0,\sigma_0^2\right).
\label{eq:t_cmb}
\end{equation}

On the contrary, the noise is Gaussian, uncorrelated, but anisotropic. Hence the noise variance depends on the position on the sky, $\sigma_{\mathrm{noise}}^2(\mathbf{x})$, but it can be calculated since the number of observations per pixel is known

\begin{equation}
T_{\mathrm{noise}}\left(\mathbf{x} \right) \sim N\left(0,\sigma_{\mathrm{noise}}^{2}\left(\mathbf{x} \right)\right).
\label{eq:t_noise}
\end{equation}

Note that the correlated WMAP noise was shown to be negligible in section 4.2 of \citet{mon08}.
The aim of the present paper is to estimate $\sigma_0^2$, comparing the obtained value to that of Gaussian simulations and checking the isotropy of the CMB. We use a new estimator but let us review first the estimator used in \citet{mon08}.

\subsection{The minimum Kolmogorov-Smirnov distance estimator}

The normalised CMB temperature is given by
\begin{equation}
u \left(\mathbf{x} \right) = \frac{T_c \left(\mathbf{x}\right)}{\sqrt{\sigma_{0}^{2}+\sigma_{\mathrm{noise}}^{2}\left(\mathbf{x}\right)}}.
\label{eq:norm_temp}
\end{equation}

According to the description given above, $u(\mathbf{x})$ has a normal distribution:

\begin{equation}
u \left(\mathbf{x} \right) \sim N(0,1).
\label{eq:norm}
\end{equation}

In order to estimate the CMB variance, we replace $\sigma_0^2$ by $s_0^2$, which we allow to vary in a sufficiently large range. In this manner we can find the value which minimizes the Kolmogorov-Smirnov distance to a normalised Gaussian. This value is our estimate for $\sigma_0^2$.
Hence we can write the minimum Kolmogorov-Smirnov estimator (hereafter MKS estimator) as:

\begin{equation}
\hat{\sigma_0^2} = s_0^2:\mathrm{min} \left|\left| \mathrm{max}\left(F_u\left(s_0^2 \right) - \Phi\right) \right|\right|,
\label{eq:mks}
\end{equation}

where $\Phi$ is the Cumulative distribution function of a standard normal and $F_u\left(s_0^2 \right)$ is the empirical distribution function for the normalised temperature.

\subsection{The unit variance estimator}

Here we propose an alternative way of estimating $\sigma_0^2$. We replace $\sigma_0^2$ by a set of values $s_0^2$ varying in a sufficiently large range and find the value where the variance of the normalised map, $\mathrm{var}\left(u\left(s_0^2\right)\right)$ is closer to unity. This value of $s_0^2$ is the estimated $\sigma_0^2$:

\begin{equation}
\hat{\sigma_0^2} = s_0^2: \mathrm{min} \left|\left| \mathrm{var}\left(u\left(s_0^2 \right)\right) -1 \right|\right|
\label{eq:uv}
\end{equation}

Let us check how this unit variance estimator (hereafter UV estimator) performs compared to the MKS estimator.

\subsection{Comparing both estimators}

Two estimators can be compared in terms of their mean squared error, which can be written as the sum of the variance and the squared bias.
For an estimator $T$ trying to estimate the true value  $\theta$ we have that the mean squared error reads:

\begin{equation}
\mathrm{m.s.e}(T)=\mathrm{E}\left[\left(T - \theta  \right)^2\right]=\mathrm{var}(T)+b^2_{T}(\theta),
\label{eq:mse}
\end{equation}
where $b_{T}(\theta)$ is the bias:

\begin{equation}
b_{T}(\theta)=\mathrm{E}\left[T \right]-\theta.
\label{eq:bias}
\end{equation}

\begin{figure}
  \begin{center}
    \includegraphics[width=84mm]{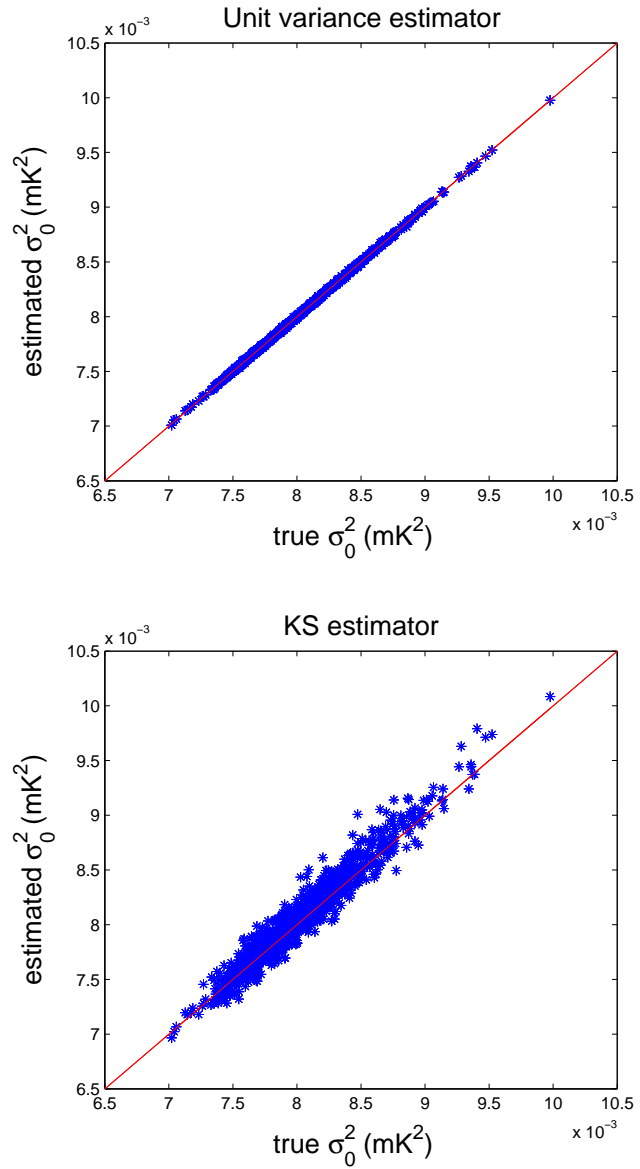}
  \end{center}
  \caption{Estimated versus true dispersion using the UV estimator (top) and the 
  MKS estimator (bottom) for $10^3$ Gaussian simulations. The kq75 mask is applied in both cases.}
  \label{fig2}
\end{figure}

Since we only have one sky we estimate the mean squared error, variance and bias using simulations. Each simulation has a different  $\sigma_0^2$, which can be calculated before adding the noise. The estimation of this quantity, $\hat{\sigma_0^2}$, is then obtained applying the estimator to the simulation including the noise. Hence we take $T=\hat{\sigma_0^2}-\sigma_0^2$ and $\theta=0$. Note that $T$ corresponds to the distance of each dot to the red line in Figure \ref{fig2}.
We apply the MKS and UV estimators to $10^3$ simulations using the kq75 mask. 
The results are shown in Table \ref{tab:bias}. The UV estimator performs better due to its lower bias and variance.

\begin{table*}
 \begin{center} \caption{Estimated mean squared error, variance and bias of $T=\hat{\sigma_0^2}-\sigma_0^2$ for the Unit Variance (UV) and Minimum Kolmogorov-Smirnov (MKS) estimators. $10^3$ simulations have been used, applying the kq75 mask.}
   \label{tab:bias}
  \begin{tabular}{c c c c}
  \hline
  Estimator $T=\hat{\sigma_0^2}-\sigma_0^2$ &  Mean Squared Error ($mK^4$) & Bias ($mK^2$) & Variance of the estimator ($mK^4$) \\
  \hline 
  UV  & $8.72 \times 10^{-11}$ & $ 3.43 \times 10^{-7}$ & $8.71 \times 10^{-11}$ \\  
  \hline
  MKS & $1.23 \times 10^{-8}$ & $-3.19 \times 10^{-5}$ & $1.12 \times 10^{-8}$ \\    
  \hline 
  \end{tabular}
 \end{center}
\end{table*}

Since we have found that the UV estimator performs better than the MKS one used in \citet{mon08}, let us repeat the variance analysis with the new tool.

\section{The Analysis}

Applying the UV estimator to the WMAP foreground cleaned VW combined map we find $\hat{\sigma_0^2}=7.18\times 10^{-3}mK^2$ which compared to $10^4$ simulations gives a $p$-value of $0.31\%$. The kq75 exclusion mask was used in this analysis. This $p$-value is much lower than the $1.2\%$ one found in \citet{mon08} using the 3--year QVW combined map, masked with the 3--year kp0 mask. 

The significance of the anomaly is high. No a posteriori choices have been made since the WMAP foreground cleaned VW combined map and the kq75 mask are recommended by the WMAP team for such analyses.

However the cause of the anomaly is unknown. In the following subsections we perform a number of follow-up tests to further investigate its possible origin, but these a posteriori tests are not valid to establish the significance of the anomaly. To make this clear, we will not consider the lower tail probabilities of the follow-up tests as $p$-values.

Let us start applying different exclusion masks to analyse the possible cause of the anomaly.

\subsection{Applying different exclusion masks}

\begin{figure}
  \begin{center}
    \includegraphics[width=84mm]{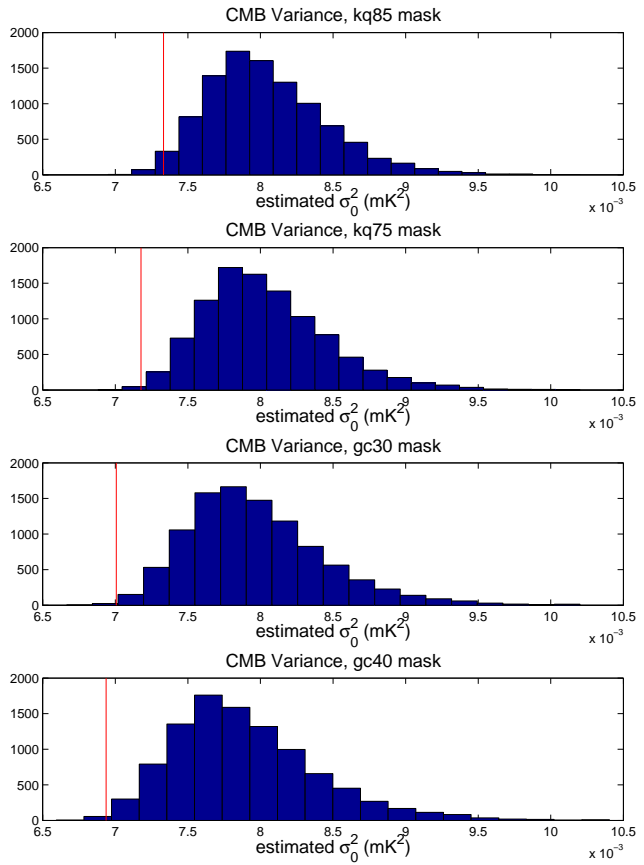}
  \end{center}
  \caption{Histograms of the variance of $10^4$ simulations, using 4 different masks, 
    namely kq85, kq75, gc30 and gc40. The red vertical line represents the value obtained for the data.}
  \label{fig3}
\end{figure}

In Table \ref{tab:signif} we list the results for the  kq85, kp0, kq75, gc30 and gc40 masks in increasing order of masked pixels (see section 2 for the definition of the masks). The corresponding histograms are shown in Figure \ref{fig3}.
The three year kp0 exclusion mask used in \citet{mon08} excluded less pixels than the kq75 mask used here and that does change in some extent the significance as can be seen in Table \ref{tab:signif}. 
Note that the lower tail probability decreases as the mask grows reaching its minimum at the gc30 mask and increases slightly for the gc40 mask.

\begin{table*}
 \begin{center} 
   \caption{Estimated CMB Variance from the data (third column) using 5 different masks, 
    namely kq85, kp0, kq75, gc30 and gc40. The sky fraction admitted by each mask at $n_{side}=256$ is shown in the second column, and the lower tail probability of the data calculated from $10^4$ simulations is listed in the right column.}
   \label{tab:signif}
  \begin{tabular}{c c c c}
  \hline
  Mask & Sky Fraction Admitted (\%) & Estimated Variance ($mK^2$) & lower tail probability  \\
  \hline
  kq85 &          81                & $ 7.33\times 10^{-3}$        & $1.57\%$   \\  
  \hline 
  kp0  &          76                & $ 7.26\times 10^{-2}$        & $0.70\%$   \\  
  \hline 
  kq75 &          71                & $ 7.18\times 10^{-3}$        & $0.31\%$   \\  
  \hline 
  gc30 &          48                & $ 7.01\times 10^{-3}$        & $0.23\%$   \\  
  \hline 
  gc40 &          34                & $ 6.94\times 10^{-3}$        & $0.41\%$   \\  
  \hline 
%  gp33 &          33                & $ 7.79\times 10^{-3}$        & $34.6\%$   \\  
%  \hline 

  \end{tabular}
 \end{center}
\end{table*}

The estimation of the variance and hence the variation of the lower tail probability is affected by the efficiency of the estimator.
If we assume that the low variance was intrinsic to the all-sky CMB data,  we would expect the lower tail probability to have its lowest value for the smallest mask (i.e. kq85) since the variance of the estimator is higher when less pixels are taken into account. 
Applying the UV estimator to $10^3$ simulations using the kq75 and gc40 masks, we find that the m.s.e of the estimator is $36\%$ higher for the gc40 mask respect to the kq75 one. The estimation is worse since less pixels are used. This has to be the reason for the higher lower tail probability for the gc40 mask compared to the gc30 one. Note that the histograms in Figure \ref{fig3} are broader for bigger masks due to this effect.

However increasing the admitted fraction of the sky from the gc30 to the kq85 mask the lower tail probability grows instead of decreasing, hence this can not be due to estimator efficiency. This trend rather suggests a higher variance in the data near the Galactic plane, which could perhaps be due to Galactic foregrounds.
Before further investigating this result, let us analyse the ecliptic hemispheres separately as was done in \citet{mon08}.

\subsection{Hemispherical analysis}

\citet{mon08} found that the significance of the low variance was higher in the northern ecliptic hemisphere. Therefore we also analyse both ecliptic hemispheres separately, finding the same result but again with increased significance. Applying the kq85 mask, the northern ecliptic hemisphere shows a lower tail probability of $0.01\%$ while the southern one is compatible with the simulations (lower tail probability $36.23\%$). A similar result is found analysing Galactic, instead of ecliptic hemispheres. We find a lower tail probability of $0.03\%$ for the Galactic north and $36.82\%$ for the Galactic south (see Table \ref{tab:signif2}).

\subsection{Localising the anomalous region }

\begin{figure}
  \begin{center}
    \includegraphics[width=4cm]{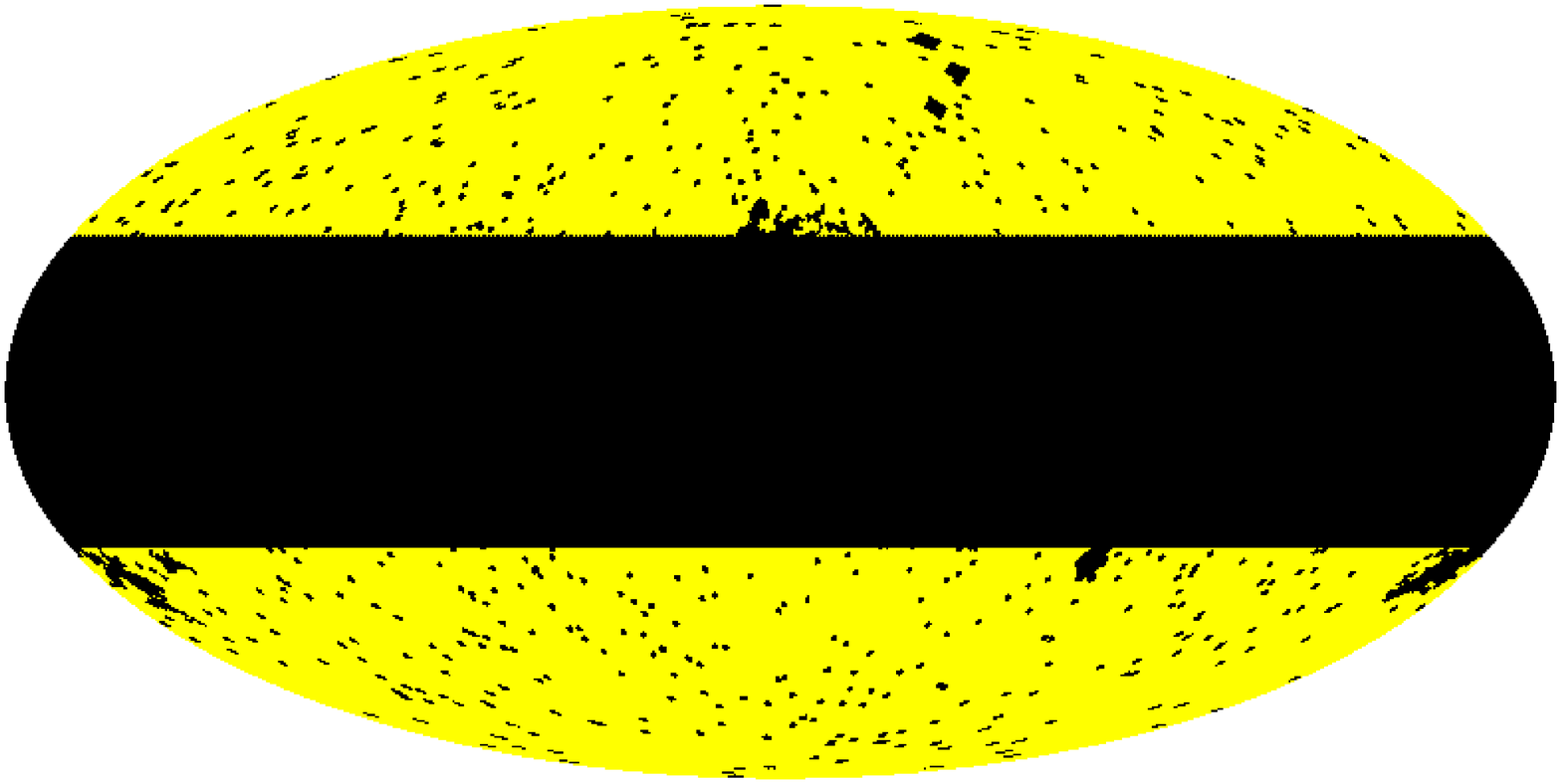}
    \includegraphics[width=4cm]{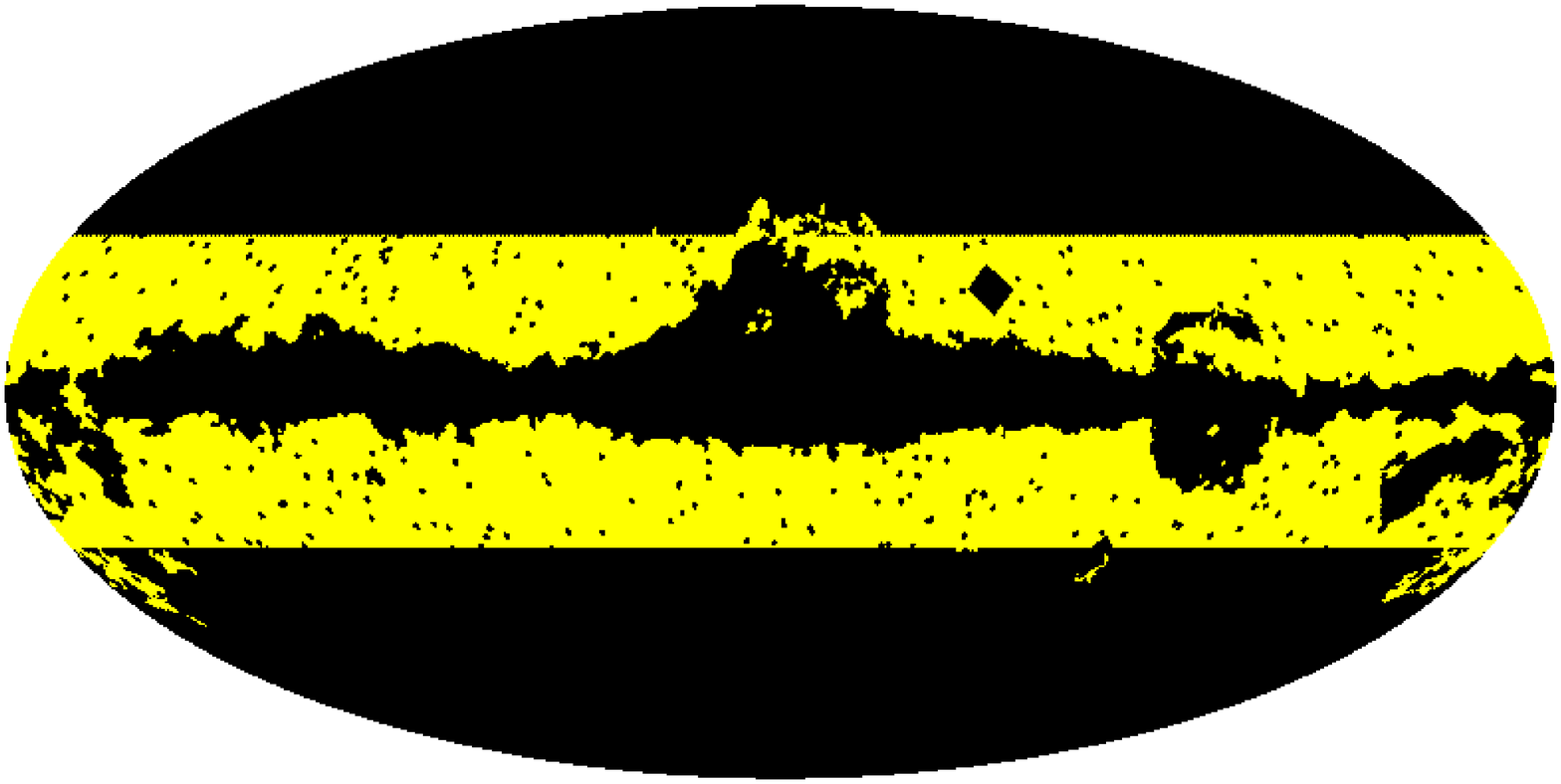}
    \includegraphics[width=4cm]{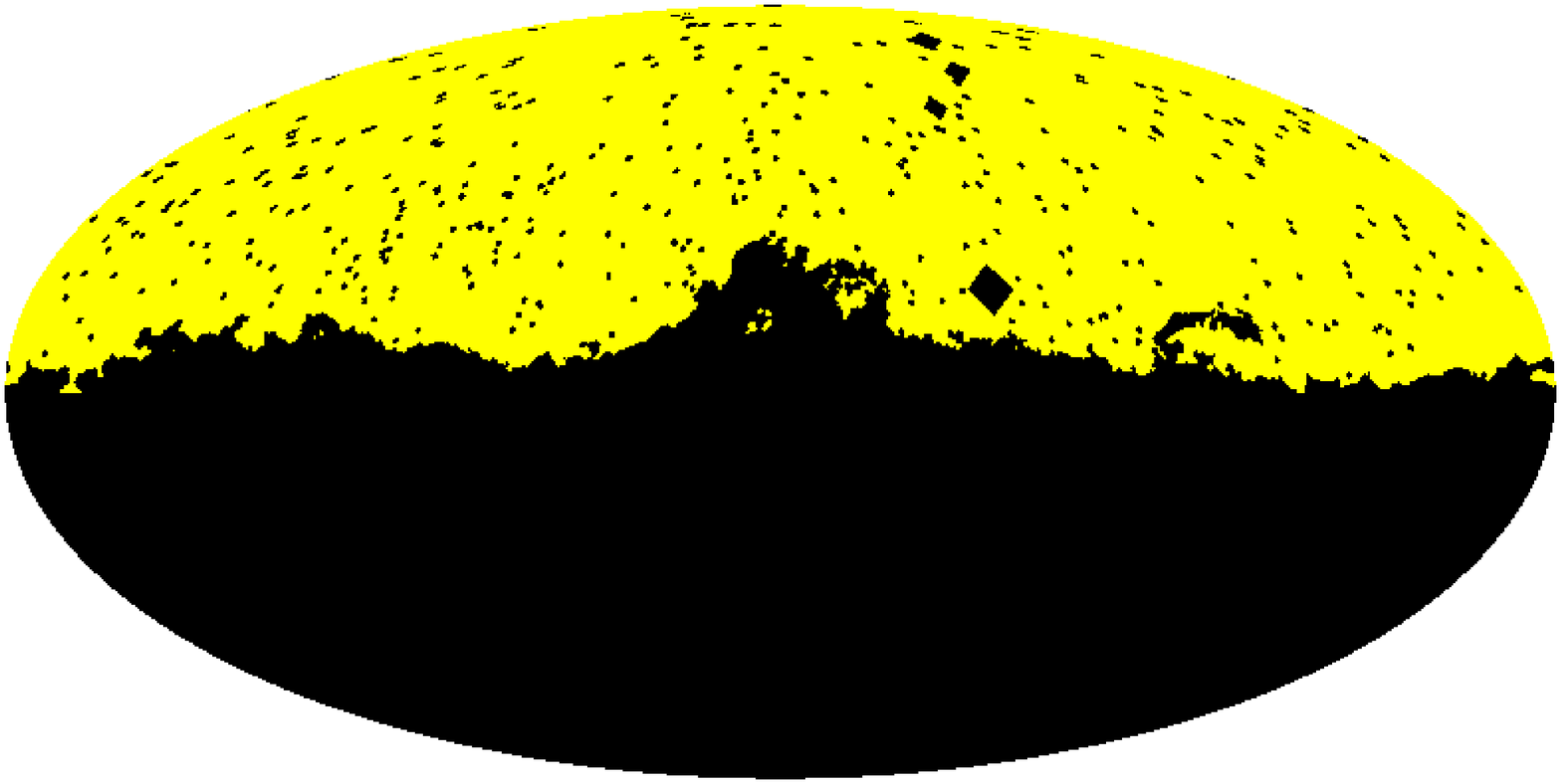}
    \includegraphics[width=4cm]{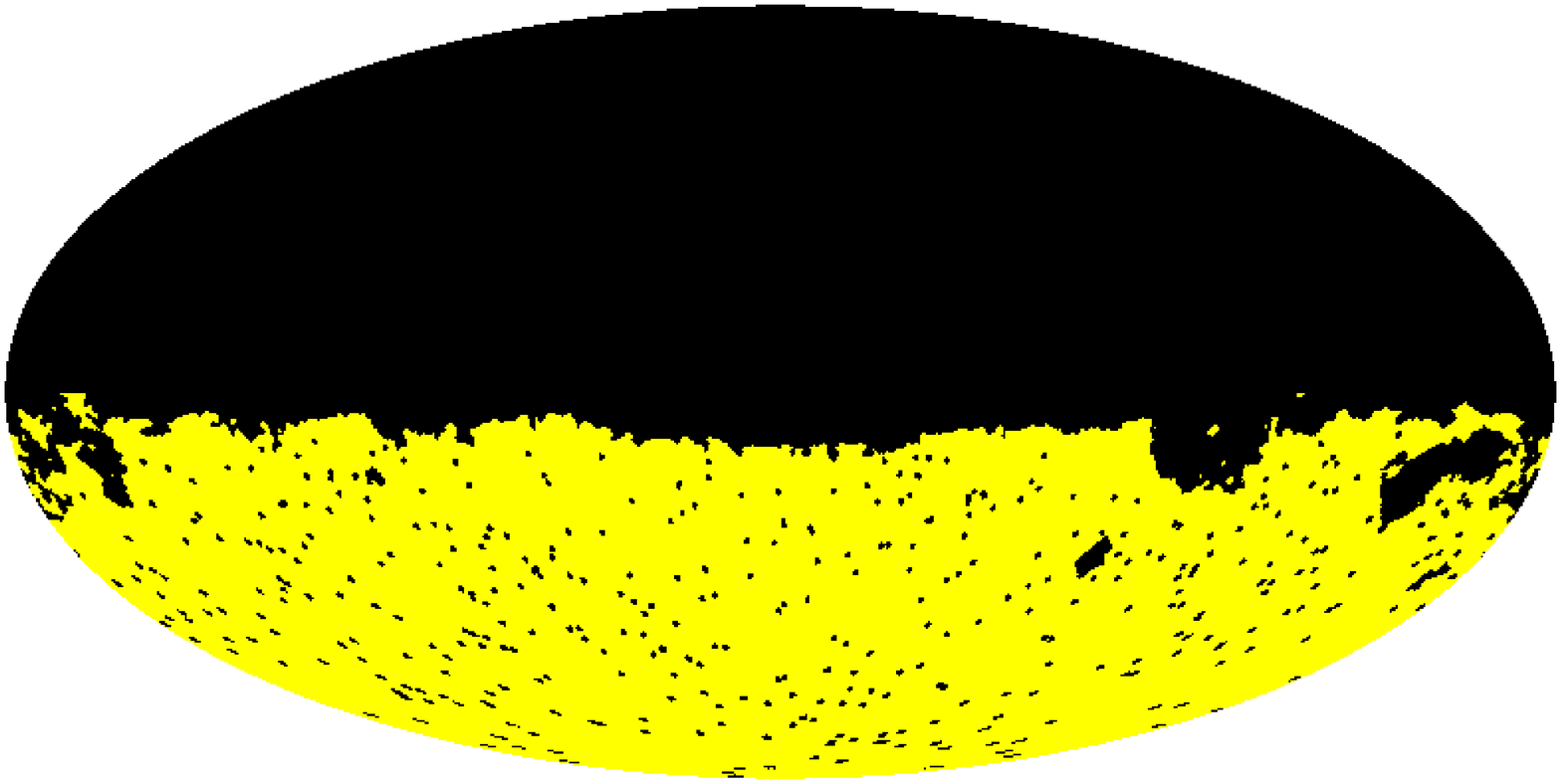}
    \includegraphics[width=4cm]{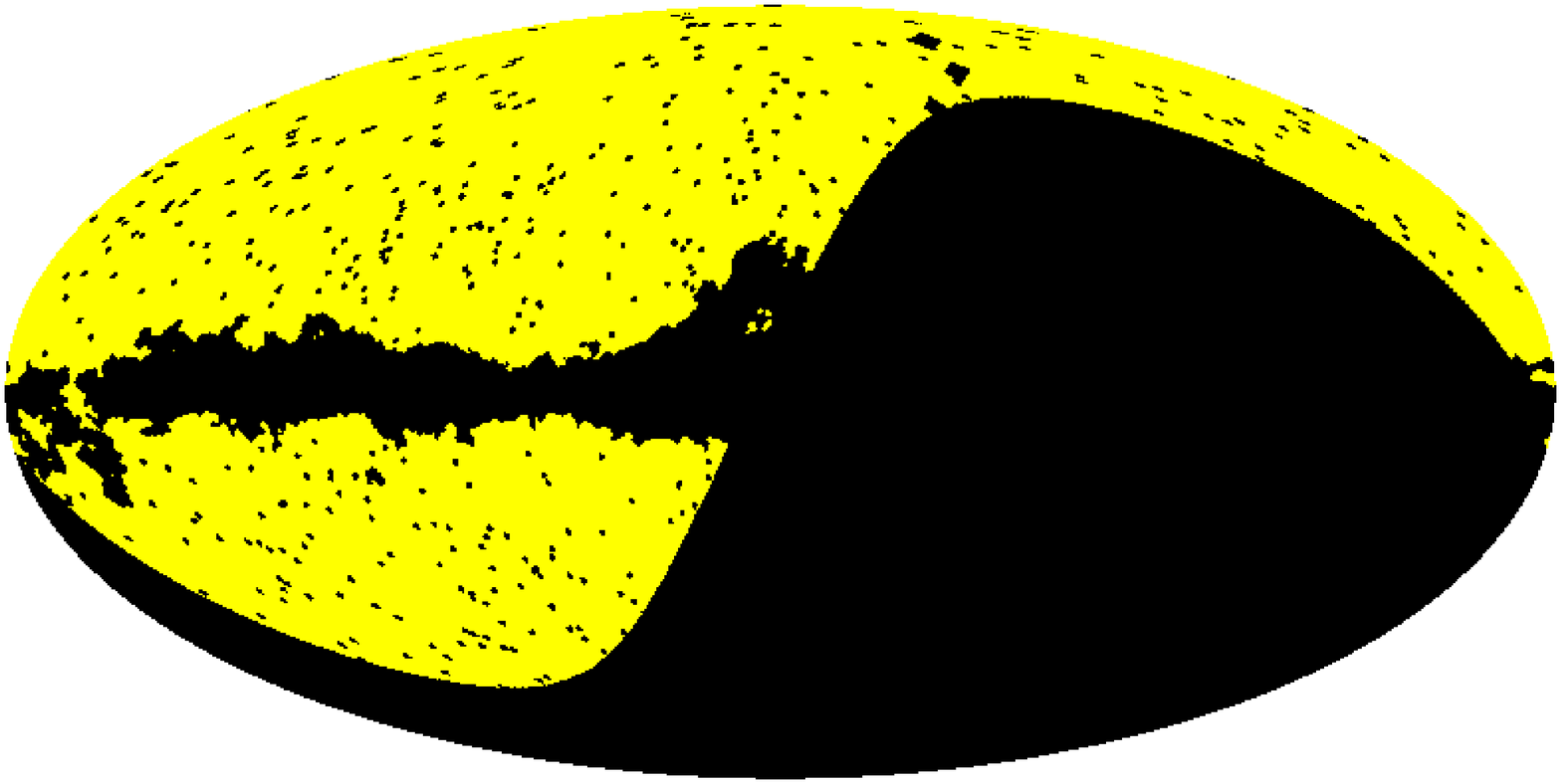}
    \includegraphics[width=4cm]{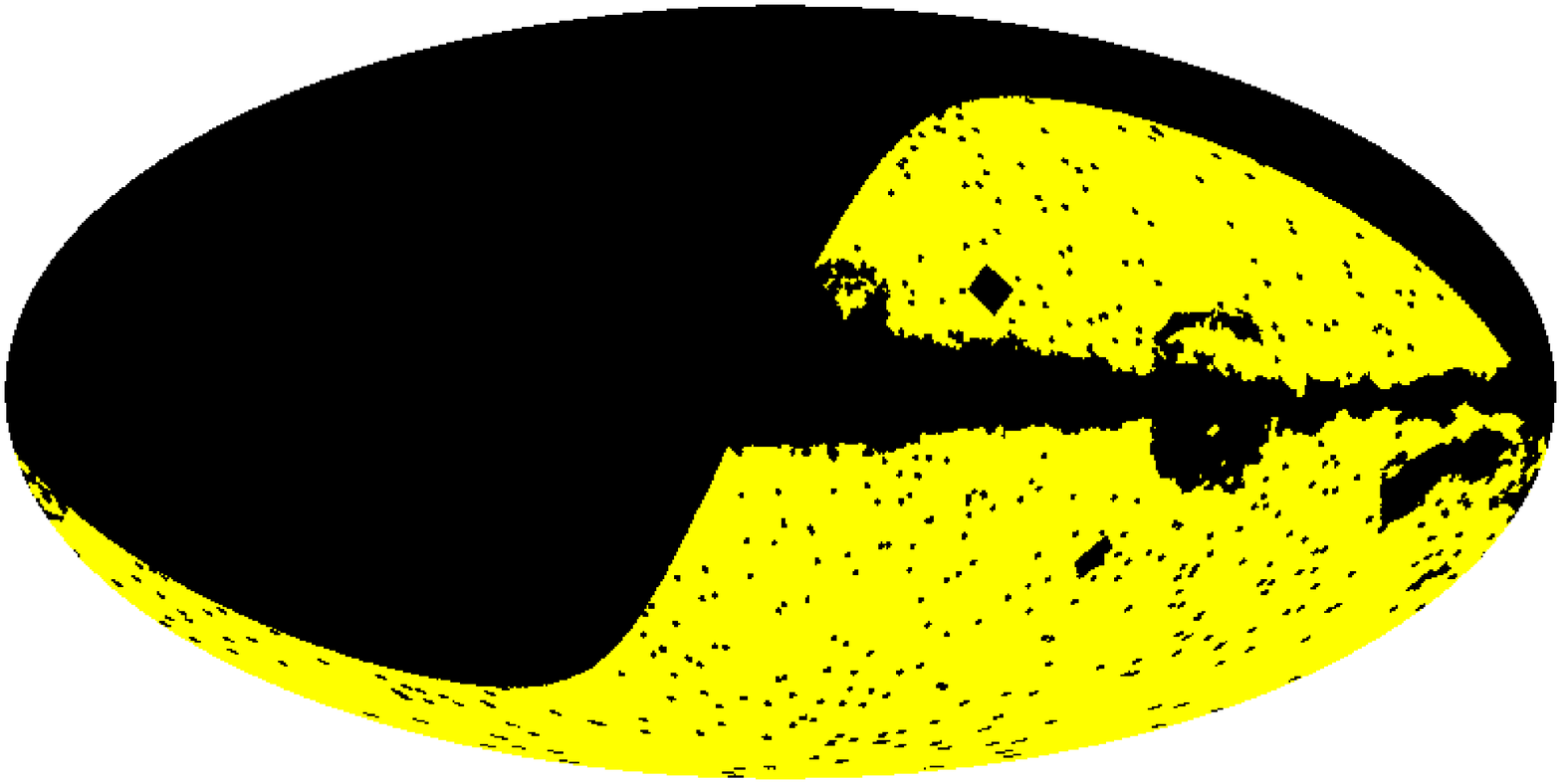}
    \includegraphics[width=4cm]{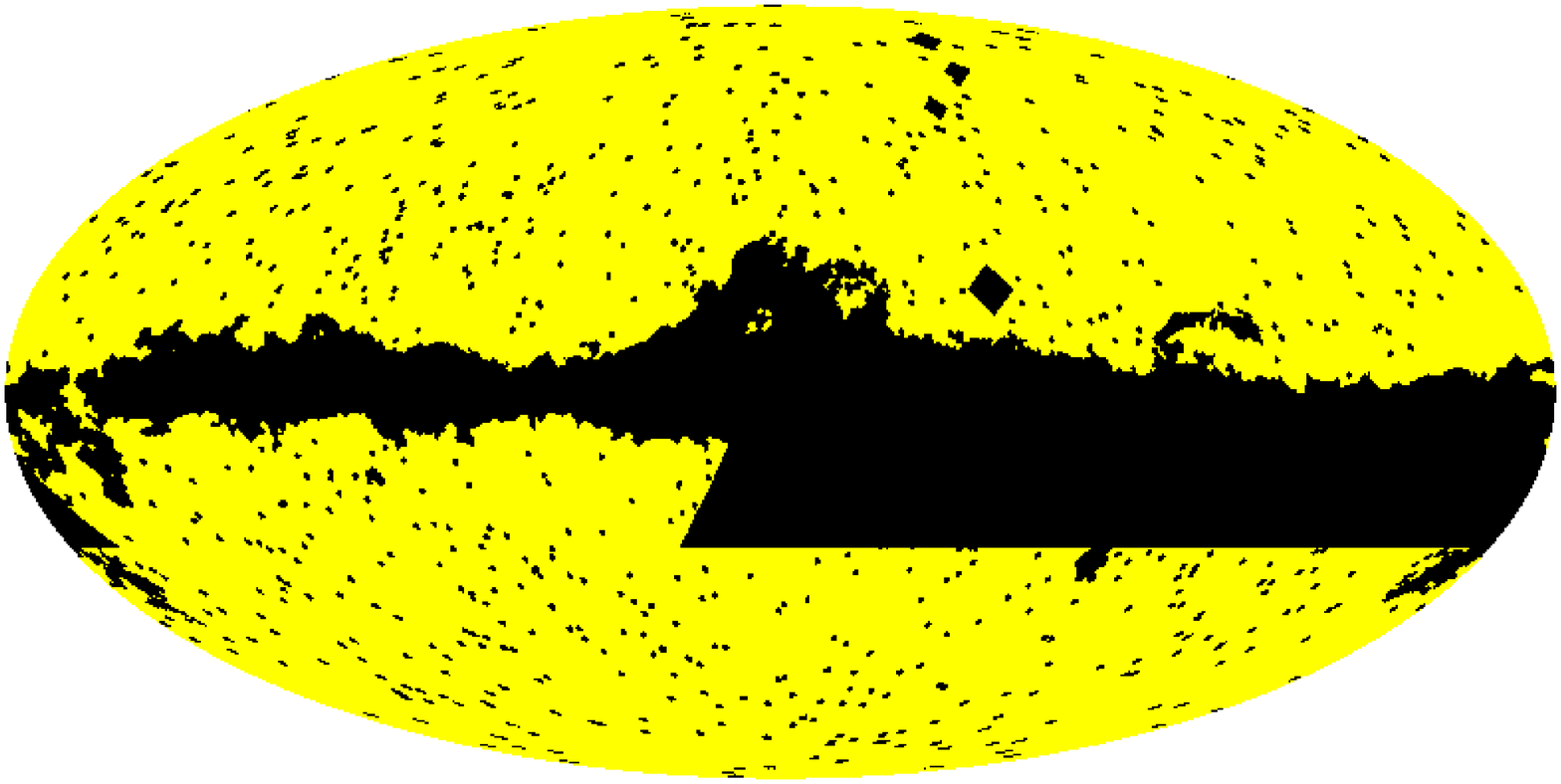}
    \includegraphics[width=4cm]{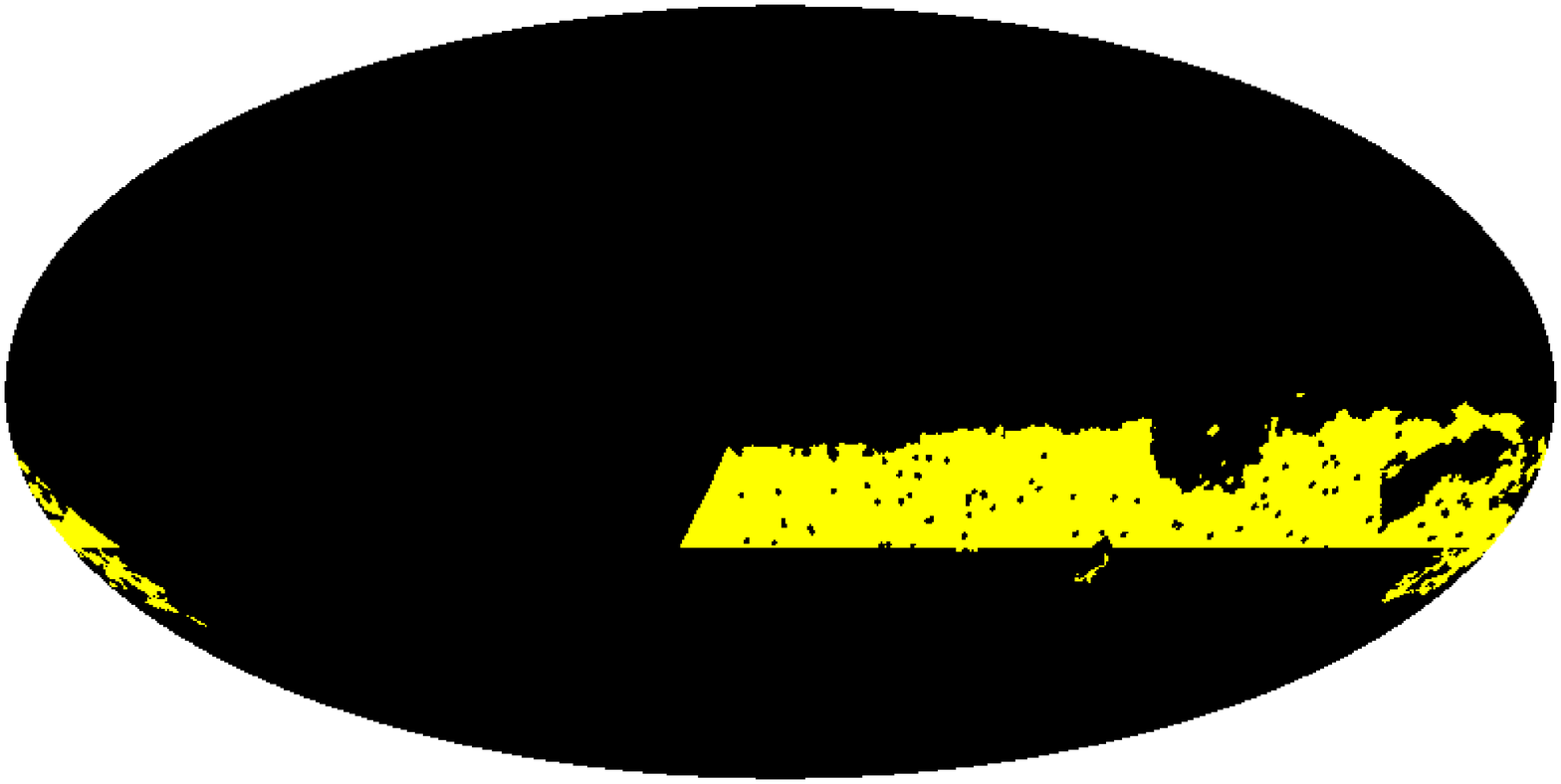}
  \end{center}
    \caption{Mollweide projections of the eight exclusion masks used in this section. From top to bottom and left to right we have the gc30, gp33, Galactic north (galN), Galactic south (galS), ecliptic north (eclN), ecliptic south (eclS), kq80 and gp10 masks. The excluded regions are plotted in black.}
   \label{fig4}
\end{figure}

The most significant lower tail probability in Table \ref{tab:signif} is obtained for the gc30 case. Therefore the pixels excluded by the gc30 mask seem to have a high variance. In order to study these pixels we build a complementary mask which admits only the pixels between the kq85 mask and the gc30 one (i.e. all the pixels not excluded by the kq85 mask but excluded by the gc30 one, see Figure \ref{fig4}). 
We call this mask gp33 since the admitted fraction of the sky is about $33\%$ and is located near the Galactic plane.
The first two lines in Table \ref{tab:signif2} show the result obtained for the gp33 mask compared to the gc30 one. The gp33 result is compatible with the simulations and completely different from those found in Table \ref{tab:signif}. In the previous subsection we analysed separately Galactic and ecliptic hemispheres. The results were similar to those obtained when studying the complementary regions allowed by the gp33 and gc30 masks as we show in rows 3 to 6 in Table \ref{tab:signif2}.

\begin{table*}
 \begin{center} 
   \caption{Variance of the data (third column) using 4 different pairs of complementary masks, 
    namely Galactic north (galN) vs. Galactic south (galS), ecliptic north (eclN) vs. ecliptic south (eclS), gc30 vs gp33 and kq80 vs gp10 (see text for detailed description of these masks).
 The sky fraction admitted by each mask at $n_{side}=256$ is shown in the second column, and the lower tail probability of the data calculated from $10^4$ simulations is listed in the right column.}
   \label{tab:signif2}
  \begin{tabular}{c c c c}
  \hline
  Mask & Sky Fraction Admitted (\%) & Estimated Variance ($mK^2$) & lower tail probability  \\
  \hline 
  gc30 &          48                & $ 7.01\times 10^{-3}$        & $0.23\%$   \\  
  gp33 &          33                & $ 7.79\times 10^{-3}$        & $38.07\%$   \\  
  \hline 
  galN &          40                & $ 6.84\times 10^{-3}$        & $0.03\%$   \\  
  galS &          41                & $ 7.81\times 10^{-3}$        & $36.82\%$   \\  
  \hline 
  eclN &          41                & $ 6.86\times 10^{-3}$        & $0.01\%$   \\  
  eclS &          40                & $ 7.81\times 10^{-3}$        & $36.23\%$   \\  
  \hline
  kq80 &          72                & $ 7.04\times 10^{-3}$        & $0.03\%$   \\  
  gp10 &           9                & $ 9.39\times 10^{-3}$        & $95.58\%$   \\  
  \hline

  \end{tabular}
 \end{center}
\end{table*}

\begin{figure}
  \begin{center}
    \includegraphics[width=84mm]{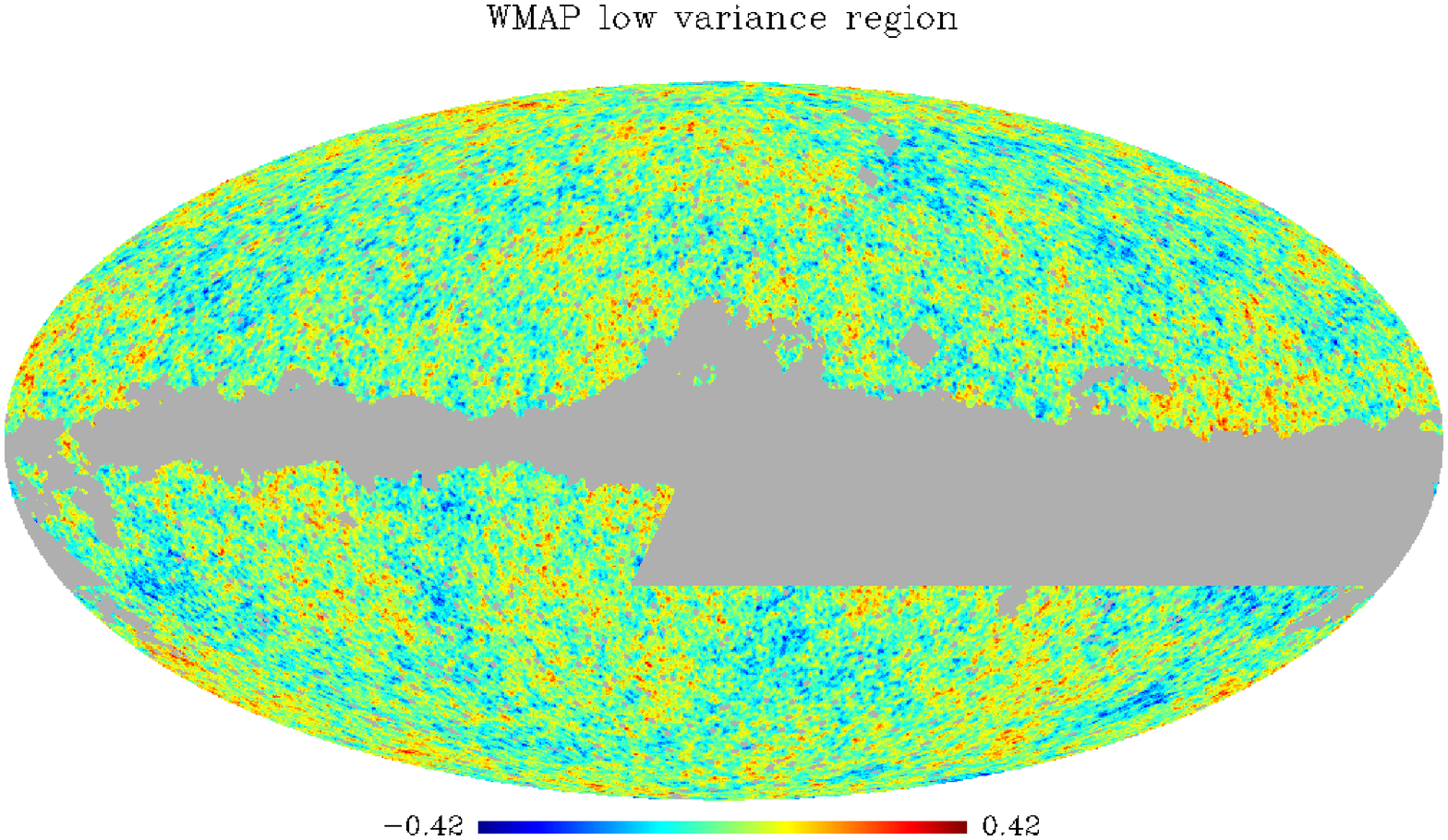}
    \includegraphics[width=84mm]{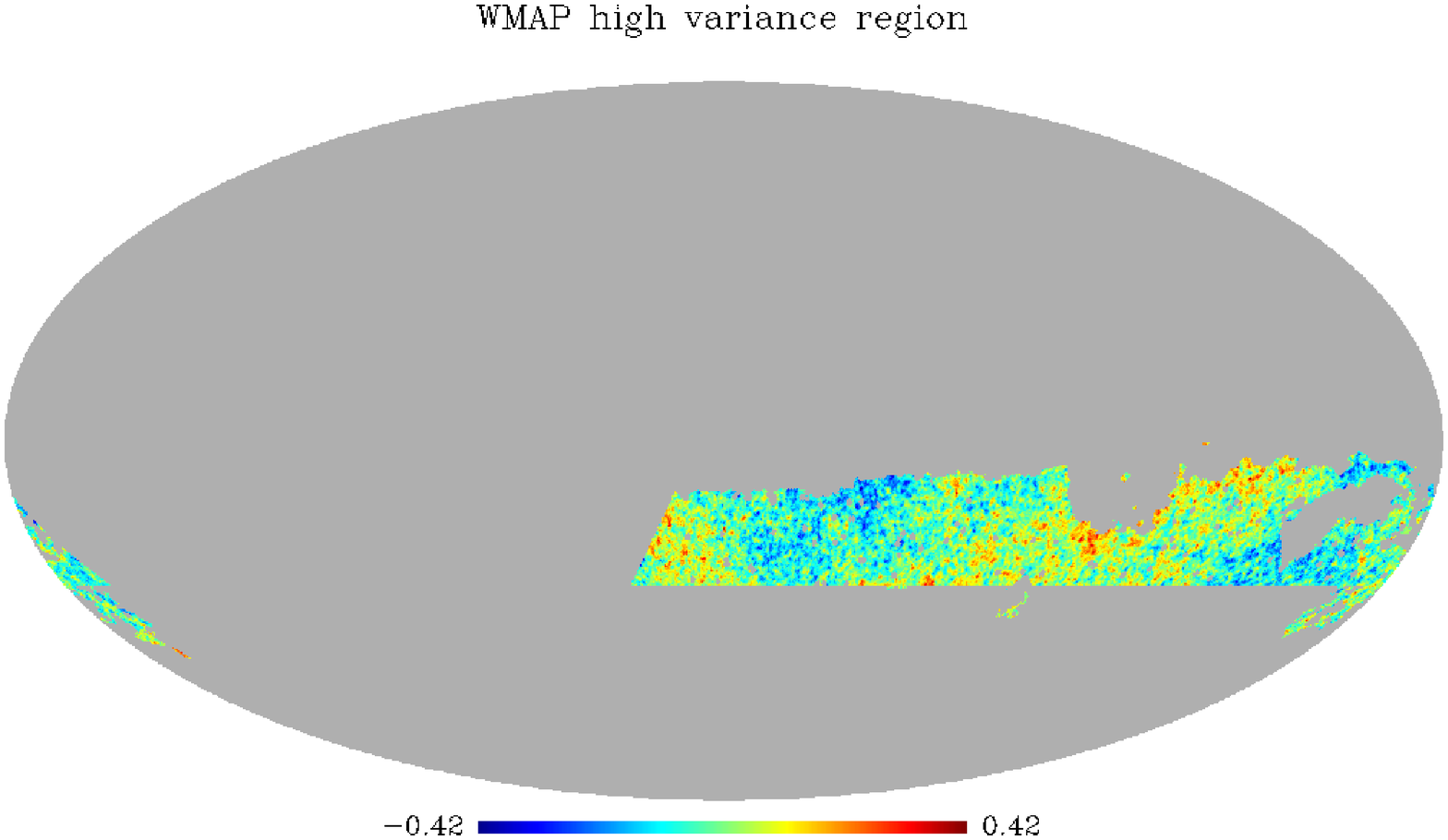}
  \end{center}
  \caption{Mollweide projection of the kq80 (top) and gp10 (bottom) masks applied to the WMAP combined VW map. The excluded regions are shown in grey.}
  \label{fig5}
\end{figure}

\begin{figure}
  \begin{center}
    \includegraphics[width=84mm]{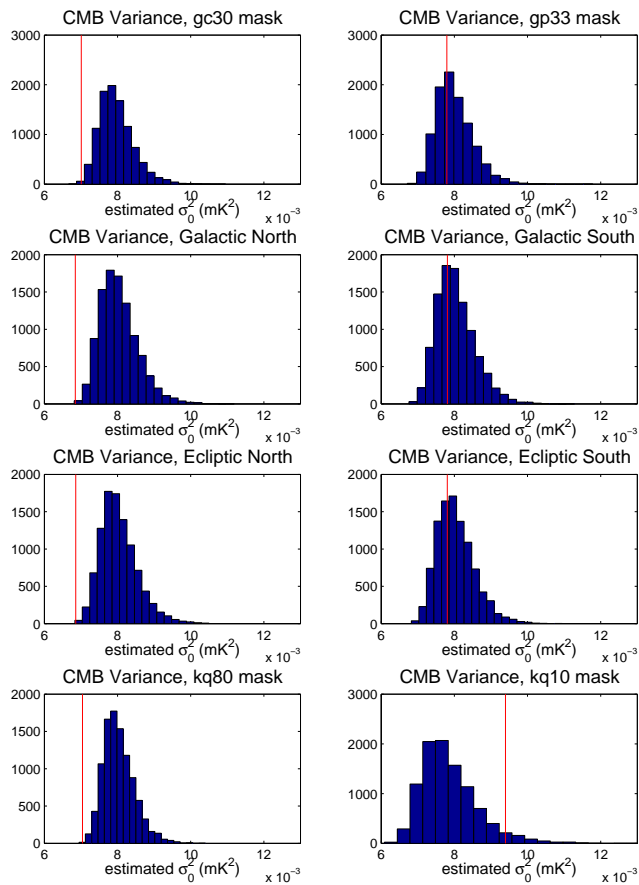}
  \end{center}
  \caption{Histograms of the variance of $10^4$ simulations, using 4 pairs of complementary masks. Each pair is plotted in a different row. From top to bottom we show gc30 versus gp33 mask, Galactic north versus Galactic south, ecliptic north versus ecliptic south and kq80 versus gp10 mask. The red vertical lines represent the values obtained for the data.}
  \label{fig6}
\end{figure}

Putting these three results together, it is clear that the intersection of the three high-variance regions (Galactic and ecliptic south, and gp33) could have an even higher variance. This intersection, i.e. the region of the sky admitted by all three masks, represents about $10\%$ of the sky. In order to study this region we build the gp10 mask admitting this small fraction of the sky and its complementary mask which we call kq80. In Figure \ref{fig5} we show the Mollweide projection of these two masks applied to the WMAP VW combined map.
The estimated variances for these masks are listed in the last two lines of Table \ref{tab:signif2}. The gp10 region shows an anomalously high variance, whereas the kq80 region presents an anomalously low variance. The histograms corresponding to Table \ref{tab:signif2} are plotted in Figure \ref{fig6}.

The relative difference of estimated variances between the WMAP gp10 and kq80 regions, is bigger than in $99.64\%$ of the simulations.
Therefore, the different variance between the region near the Galactic plane  and the rest of the sky seems to be violating the cosmological principle of isotropy and could be the cause of the anomalous variance. Isotropy is one of the key assumptions made to estimate the angular power spectrum \citep{nol09} and to generate the simulations. 
Let us recall the process we follow in this analysis. We use the best fit $C_\ell$ estimated from the data to generate Gaussian and isotropic simulations, and then we compare the variance of the simulations again to the data.

Since the kq85 mask is used for estimating the $C_\ell$ \citep{nol09} and there is an excess of power in the gp10 region, the isotropic simulations will have an excess of variance in the complementary kq80 region when compared to the data.

Hence the anomaly could not be due to an isotropic low variance of the data over the whole sky, but rather to an anisotropic power distribution. 

The high variance in the Galactic plane region is a clear hint for Galactic foreground contamination, since adding any two independent emissions ends up in an increase of the variance. 
However we would need some other support for the Galactic foreground hypothesis, since the anisotropic variance distribution could be casual. Contaminating Galactic foregrounds such as synchrotron, free-free and Galactic dust emission show a strong frequency dependence. Therefore, a frequency dependence of the variance in the gp10 region would be proving the foreground hypothesis right.

\subsection{Individual frequency bands }

Here we perform the analysis on the Q (41 GHz), V (61 GHz) and W (94 GHz) frequency bands separately. We compare the estimated variance to $10^3$ simulations for each band. 

We find a slight frequency dependence of the lower tail probability when using the gp10 mask but the significance remains almost unchanged using the kq80 one. The lower tail probabilities for the Q, V and W bands are $0.02\%$, $0.01\%$ and  $0.01\%$ respectively, using the kq80 mask; and  $96.7\%$, $95.4\%$ and  $95.7\%$ applying the gp10 mask. 

\begin{figure}
  \begin{center}
    \includegraphics[width=84mm]{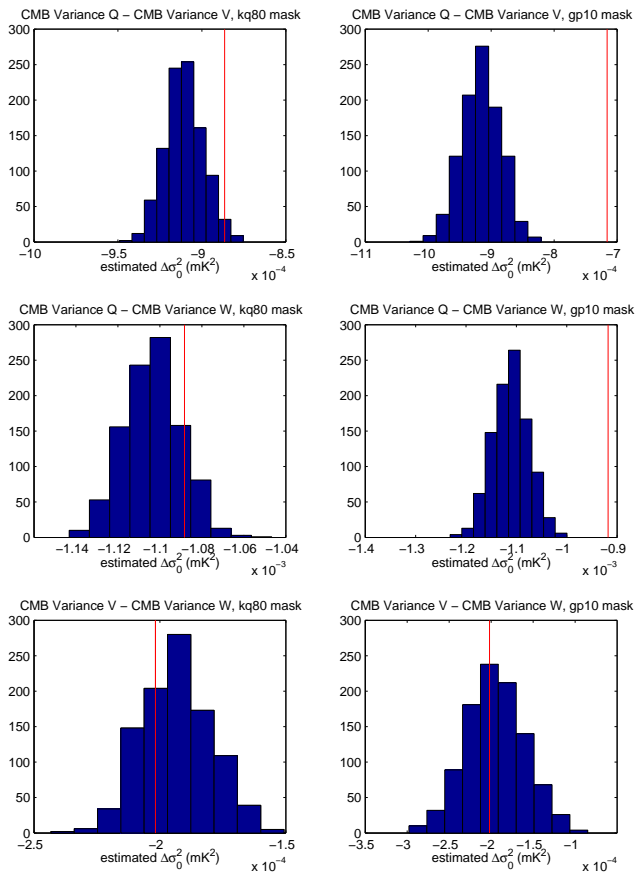}
  \end{center}
  \caption{Histograms of the differences of estimated variances of the Q and V bands (top row), Q and W bands (middle row), V and W bands (bottom row). The kq80 mask has been used for panels of the left column and the gp10 one for the right column. $10^3$ simulations have been used for each band. The data are represented by a red vertical line.}
  \label{fig7}
\end{figure}

In order to check whether the frequency dependence of the variance is significant, we calculate the difference of the estimated variance between different bands, and compare their histograms to the data in Figure \ref{fig7}.

The gp10 variance differences including the Q band data, show a very high frequency dependende. This result indicates the presence of some foreground residuals near the Galactic plane in the Q band. The V and W bands which are the cosmological bands, do not show any anomalous frequency dependence, but the Q band result and the anomalously high variance make this region highly suspicious of being contaminated. The combination of dust and synchrotron or free-free emission could perhaps give a flat frequency dependence at the V and W frequencies. Moreover an artifact of the foreground cleaning process could be affecting the pixels in this region.

The excess of variance near the Galactic plane could be related to other, previously reported anomalies.
For instance, many authors (see introduction) reported an anomalous quadrupole-octopole alignment in a direction near the Galactic plane.

\subsection{Subtracting Quadrupole and Octopole}

\begin{figure}
  \begin{center}
    \includegraphics[width=84mm]{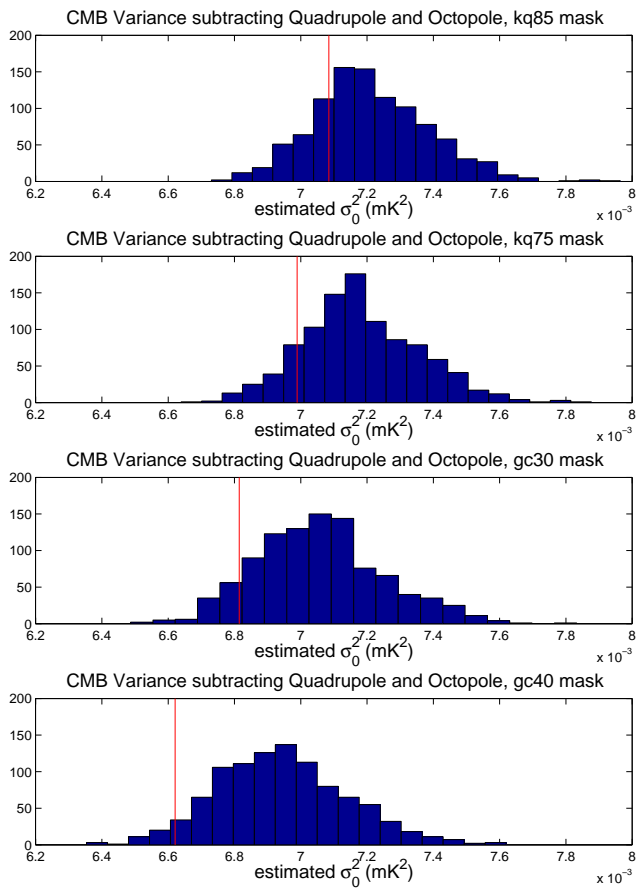}
  \end{center}
  \caption{Histogram of the variance removing the quadrupole and octopole from data and simulations. Four different masks are used. From top to bottom we apply the kq85, kq75, gc30 and gc40 masks.}
  \label{fig8}
\end{figure}

\begin{figure}
  \begin{center}
    \includegraphics[width=84mm]{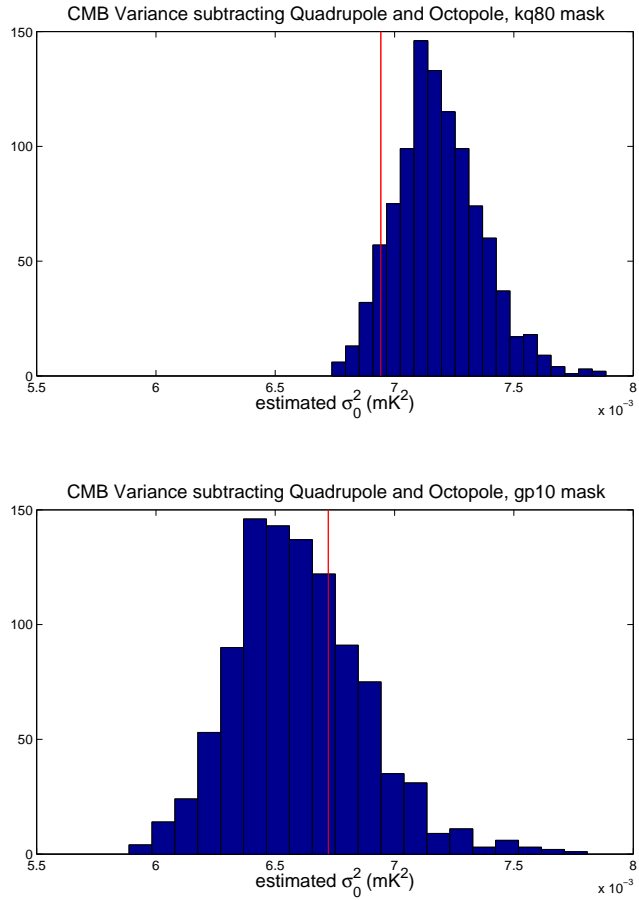}
  \end{center}
  \caption{Histogram of the variance removing the quadrupole and octopole from data and simulations. The kq80 (top) and gp10 (bottom) masks are used.}
  \label{fig9}
\end{figure}

Let us study a possible relation  between the low variance and the anomalous quadrupole-octopole alignment.
In order to check the possible influence of these multipoles in our results, we remove the quadrupole and octopole from data and simulations, and repeat the analysis. The results plotted in Figure \ref{fig8} and Figure \ref{fig9} show that the data are now fully compatible with the simulations. The link between both anomalies is clear and they could have a common cause.
Note that Figure \ref{fig8} still reveals a slight dependence of the variance with the mask, i.e. the less pixels in the Galactic region are admitted, the lower is the variance of the data as compared to simulations. This could hint again to the presence of residual Galactic foreground contamination, but the variation is not statistically significant.
Some authors, such as \citet{slo04} or \citet{abr10}, pointed out foreground residuals as the cause of the quadrupole-octopole alignment. The results in this section show us that if there are Galactic foregrounds in the gp10 region, they mainly affect the quadrupole and octopole.


\subsection{WMAP 7 year data}

At the same time this paper was written, the WMAP team released a new version of the WMAP data \citep{jar10}. 
The data and masks given in the new release are very similar to the 5-year ones.
We estimated the CMB variance for the 7-year data using the 7-year kq75 and kq85 masks obtaining $7.17\times 10^{-3}mK^2$ and $7.27 \times 10^{-3} mK^2$ respectively, which differ less than $0.1\%$ and $0.9\%$ from the 5-year results listed in Table \ref{tab:signif}. Similar differences are found using the other masks. 
The lower instrumental noise of the 7-year data has a negligible effect in our variance estimation.
The estimated variance for the 5-year data using the 7-year kq85 mask is $7.26 \times 10^{-3} mK^2$ which is a variation of $0.1\%$ with respect to the 7-year data result.
Hence our results and conclusions remain unaffected by these small changes.

\section{Conclusions}

In this paper we compare the variance of the WMAP data to Gaussian simulations generated from the best fit angular power spectrum estimated from the data. Isotropy and Gaussianity are assumed to generate the simulations. \citet{mon08} already performed this comparison for the 3--year WMAP data, using the MKS variance estimator. They found that the variance of the data was lower than $98.8\%$ of the simulations. 

Here we apply the UV estimator to the 5--year WMAP data. The UV estimator performs better than the MKS one due to a lower estimator variance and bias. We find again a low CMB variance with increased significance. The $p$-value for the kq75 mask is $0.31\%$. Performing the same analysis applying different exclusion masks we find a strong dependence of the variance with the used exclusion mask. The region near the Galactic plane shows a higher variance and compatibility with simulations, while the complementary region near the Galactic poles shows an anomalously low variance. We further localise the high-variance region, finding that the excess in variance is caused by about $10\%$ of the sky. This region, which we call the gp10 region, is located near the Galactic plane and shows an anomalously high variance (lower tail probability = $95.58\%$), whereas its complementary region (kq80 region) has an anomalously low variance (lower tail probability = $0.03\%$).

The relative difference of estimated variances between the WMAP gp10 and kq80 regions, is bigger than in $99.64\%$ of the simulations.
Hence, this anisotropic distribution of the variance could be the cause of the observed anomaly. The angular power spectrum has been estimated applying the kq85 mask, admitting the high-variance region near the Galactic plane. Since isotropy is assumed, this high power is distributed over the whole sky when generating simulations. Comparing this \emph{high-variance} simulations to the low variance region of the sky, they turn out to be incompatible.

Studying the relation of the low variance anomaly to the quadrupole-octopole alignment, we find that removing the quadrupole and octopole from data and simulations, the variance anomaly disappears. This link hints for a common cause of both anomalies.

There are several possible causes for both anomalies, namely unknown systematic effects, fortuitous alignment of quadrupole and octopole, Galactic foregrounds and cosmic defects.
Any unknown systematic effect could be causing the anomaly, although there is no evidence for this hypothesis yet. A fortuitous alignment of quadrupole and octopole has been proven to be very unlikely \citep{ben10}. 
A high variance in the region near the Galactic plane is a hint for Galactic foreground contamination. Furthermore we find strong evidence for foreground contamination in the Q band high-variance region. \citet{slo04} already pointed out Galactic foregrounds as the cause of the alignment. \citet{ben10} shows that masking some small regions in the sky the alignment disappears. In particular, masking some spots in the high-variance region, eliminates any significant alignment. This shows that the alignment is not robust and could be due to Galactic foregrounds in the high-variance region. Topological defects such as textures \citep{cru07b} could also be considered as an alternative explanation.

In our opinion the most likely hypothesis given  the frequency dependence with respect to the Q band and the location of the high-variance region are Galactic foreground residuals. If this hypothesis was proven right, the cosmological parameter estimation could be affected in some extent. However, this has to be carefully analysed in future work.

\section*{Acknowledgments}

We acknowledge financial support from the Spanish MCYT project AYA2007-68058-C03-02.
PV acknowledges financial support from the Ram\'on y Cajal project. 
We acknowledge the use of the Legacy Archive for Microwave Background Data 
Analysis (LAMBDA). Support for LAMBDA is provided by the NASA Office of Space 
Science.
This work has used the software package HEALPix (Hierarchical, Equal
Area and iso-latitude pixelization of the sphere,
http://www.eso.org/science/healpix), developed by K.M. G{\'o}rski,
E. F. Hivon, B. D. Wandelt, J. Banday, F. K. Hansen and
M. Barthelmann; and the CAMB and CMBFAST software, developed by  A. Lewis and A. Challinor and 
by U. Seljak and M. Zaldarriaga respectively.

\bibliographystyle{mn2e}
\bibliography{apj-jour,bibliografia}

\begin{thebibliography}{}

\bibitem[\protect\citeauthoryear{{Abramo}, {Bernui} \& {Pereira}}{{Abramo}
  et~al.}{2009}]{abr10}
{Abramo} L.~R.,  {Bernui} A.,    {Pereira} T.~S.,  2009, JCAP, 12, 13

\bibitem[\protect\citeauthoryear{{Abramo}, {Sodr{\'e}} Jr. \&
  {Wuensche}}{{Abramo} et~al.}{2006}]{abr06}
{Abramo} L.~R.,  {Sodr{\'e}} Jr. L.,    {Wuensche} C.~A.,  2006, \prd, 74,
  083515

\bibitem[\protect\citeauthoryear{{Ayaita}, {Weber} \& {Wetterich}}{{Ayaita}
  et~al.}{2009}]{aya09}
{Ayaita} Y.,  {Weber} M.,    {Wetterich} C.,  2009, ArXiv e-prints

\bibitem[\protect\citeauthoryear{{Bennett} et~al.,}{{Bennett}
  et~al.}{2003}]{ben03}
{Bennett} C.~L.,  et~al., 2003, \apjs, 148, 1

\bibitem[\protect\citeauthoryear{{Bennett} et~al.,}{{Bennett}
  et~al.}{2010}]{ben10}
{Bennett} C.~L.,  et~al., 2010, ArXiv e-prints

\bibitem[\protect\citeauthoryear{{Bernui}}{{Bernui}}{2009}]{ber09}
{Bernui} A.,  2009, \prd, 80, 123010

\bibitem[\protect\citeauthoryear{{Bernui}, {Mota}, {Rebou{\c c}as} \&
  {Tavakol}}{{Bernui} et~al.}{2007}]{ber07}
{Bernui} A.,  {Mota} B.,  {Rebou{\c c}as} M.~J.,    {Tavakol} R.,  2007, \aap,
  464, 479

\bibitem[\protect\citeauthoryear{{Bielewicz}, {Eriksen}, {Banday}, {G{\'o}rski}
  \& {Lilje}}{{Bielewicz} et~al.}{2005}]{bie05}
{Bielewicz} P.,  {Eriksen} H.~K.,  {Banday} A.~J.,  {G{\'o}rski} K.~M.,
  {Lilje} P.~B.,  2005, \apj, 635, 750

\bibitem[\protect\citeauthoryear{{Bielewicz}, {G{\'o}rski} \&
  {Banday}}{{Bielewicz} et~al.}{2004}]{bie04}
{Bielewicz} P.,  {G{\'o}rski} K.~M.,    {Banday} A.~J.,  2004, \mnras, 355,
  1283

\bibitem[\protect\citeauthoryear{{Cay{\'o}n}}{{Cay{\'o}n}}{2010}]{cay10}
{Cay{\'o}n} L.,  2010, \mnras, 405, 1084

\bibitem[\protect\citeauthoryear{{Cay{\'o}n}, {Jin} \& {Treaster}}{{Cay{\'o}n}
  et~al.}{2005}]{cay05}
{Cay{\'o}n} L.,  {Jin} J.,    {Treaster} A.,  2005, \mnras, 362, 826

\bibitem[\protect\citeauthoryear{{Cayon}, {Martinez-Gonzalez} \&
  {Sanz}}{{Cayon} et~al.}{1991}]{cay91}
{Cayon} L.,  {Martinez-Gonzalez} E.,    {Sanz} J.~L.,  1991, \mnras, 253, 599

\bibitem[\protect\citeauthoryear{{Copi}, {Huterer}, {Schwarz} \&
  {Starkman}}{{Copi} et~al.}{2006}]{cop06}
{Copi} C.~J.,  {Huterer} D.,  {Schwarz} D.~J.,    {Starkman} G.~D.,  2006,
  \mnras, 367, 79

\bibitem[\protect\citeauthoryear{{Copi}, {Huterer}, {Schwarz} \&
  {Starkman}}{{Copi} et~al.}{2007}]{cop07}
{Copi} C.~J.,  {Huterer} D.,  {Schwarz} D.~J.,    {Starkman} G.~D.,  2007,
  \prd, 75, 023507

\bibitem[\protect\citeauthoryear{{Copi}, {Huterer} \& {Starkman}}{{Copi}
  et~al.}{2004}]{cop04}
{Copi} C.~J.,  {Huterer} D.,    {Starkman} G.~D.,  2004, \prd, 70, 043515

\bibitem[\protect\citeauthoryear{{Cruz}, {Cay{\'o}n},
  {Mart{\'{\i}}nez-Gonz{\'a}lez}, {Vielva} \& {Jin}}{{Cruz}
  et~al.}{2007a}]{cru07a}
{Cruz} M.,  {Cay{\'o}n} L.,  {Mart{\'{\i}}nez-Gonz{\'a}lez} E.,  {Vielva} P.,
   {Jin} J.,  2007a, \apj, 655, 11

\bibitem[\protect\citeauthoryear{{Cruz}, {Mart{\'{\i}}nez-Gonz{\'a}lez},
  {Vielva} \& {Cay{\'o}n}}{{Cruz} et~al.}{2005}]{cru05}
{Cruz} M.,  {Mart{\'{\i}}nez-Gonz{\'a}lez} E.,  {Vielva} P.,    {Cay{\'o}n} L.,
   2005, \mnras, 356, 29

\bibitem[\protect\citeauthoryear{{Cruz}, {Mart{\'{\i}}nez-Gonz{\'a}lez},
  {Vielva}, {Diego}, {Hobson} \& {Turok}}{{Cruz} et~al.}{2008}]{cru08}
{Cruz} M.,  {Mart{\'{\i}}nez-Gonz{\'a}lez} E.,  {Vielva} P.,  {Diego} J.~M.,
  {Hobson} M.,    {Turok} N.,  2008, \mnras, 390, 913

\bibitem[\protect\citeauthoryear{{Cruz}, {Tucci},
  {Mart{\'{\i}}nez-Gonz{\'a}lez} \& {Vielva}}{{Cruz} et~al.}{2006}]{cru06}
{Cruz} M.,  {Tucci} M.,  {Mart{\'{\i}}nez-Gonz{\'a}lez} E.,    {Vielva} P.,
  2006, \mnras, 369, 57

\bibitem[\protect\citeauthoryear{{Cruz}, {Turok}, {Vielva},
  {Mart{\'{\i}}nez-Gonz{\'a}lez} \& {Hobson}}{{Cruz} et~al.}{2007b}]{cru07b}
{Cruz} M.,  {Turok} N.,  {Vielva} P.,  {Mart{\'{\i}}nez-Gonz{\'a}lez} E.,
  {Hobson} M.,  2007b, Science, 318, 1612

\bibitem[\protect\citeauthoryear{{de Oliveira-Costa}, {Tegmark}, {Zaldarriaga}
  \& {Hamilton}}{{de Oliveira-Costa} et~al.}{2004}]{oli04}
{de Oliveira-Costa} A.,  {Tegmark} M.,  {Zaldarriaga} M.,    {Hamilton} A.,
  2004, \prd, 69, 063516

\bibitem[\protect\citeauthoryear{{Donoghue} \& {Donoghue}}{{Donoghue} \&
  {Donoghue}}{2005}]{don05}
{Donoghue} E.~P.,  {Donoghue} J.~F.,  2005, \prd, 71, 043002

\bibitem[\protect\citeauthoryear{{Eriksen}, {Banday}, {G{\'o}rski}, {Hansen} \&
  {Lilje}}{{Eriksen} et~al.}{2007}]{eri07}
{Eriksen} H.~K.,  {Banday} A.~J.,  {G{\'o}rski} K.~M.,  {Hansen} F.~K.,
  {Lilje} P.~B.,  2007, \apjl, 660, L81

\bibitem[\protect\citeauthoryear{{Eriksen}, {Banday}, {G{\'o}rski} \&
  {Lilje}}{{Eriksen} et~al.}{2005}]{eri05}
{Eriksen} H.~K.,  {Banday} A.~J.,  {G{\'o}rski} K.~M.,    {Lilje} P.~B.,  2005,
  \apj, 622, 58

\bibitem[\protect\citeauthoryear{{Eriksen}, {Hansen}, {Banday}, {G{\'o}rski} \&
  {Lilje}}{{Eriksen} et~al.}{2004}]{eri04a}
{Eriksen} H.~K.,  {Hansen} F.~K.,  {Banday} A.~J.,  {G{\'o}rski} K.~M.,
  {Lilje} P.~B.,  2004, \apj, 605, 14

\bibitem[\protect\citeauthoryear{{Eriksen}, {Novikov}, {Lilje}, {Banday} \&
  {G{\'o}rski}}{{Eriksen} et~al.}{2004}]{eri04b}
{Eriksen} H.~K.,  {Novikov} D.~I.,  {Lilje} P.~B.,  {Banday} A.~J.,
  {G{\'o}rski} K.~M.,  2004, \apj, 612, 64

\bibitem[\protect\citeauthoryear{{Frommert} \& {En{\ss}lin}}{{Frommert} \&
  {En{\ss}lin}}{2010}]{fro10}
{Frommert} M.,  {En{\ss}lin} T.~A.,  2010, \mnras, 403, 1739

\bibitem[\protect\citeauthoryear{{G{\'o}rski}, {Hivon}, {Banday}, {Wandelt},
  {Hansen}, {Reinecke} \& {Bartelmann}}{{G{\'o}rski} et~al.}{2005}]{gor05}
{G{\'o}rski} K.~M.,  {Hivon} E.,  {Banday} A.~J.,  {Wandelt} B.~D.,  {Hansen}
  F.~K.,  {Reinecke} M.,    {Bartelmann} M.,  2005, \apj, 622, 759

\bibitem[\protect\citeauthoryear{{Gruppuso} \& {Burigana}}{{Gruppuso} \&
  {Burigana}}{2009}]{gru09}
{Gruppuso} A.,  {Burigana} C.,  2009, Journal of Cosmology and Astro-Particle
  Physics, 8, 4

\bibitem[\protect\citeauthoryear{{Gurzadyan}, {Allahverdyan}, {Ghahramanyan},
  {Kashin}, {Khachatryan}, {Kocharyan}, {Kuloghlian}, {Mirzoyan}, {Poghosian}
  \& {Yegorian}}{{Gurzadyan} et~al.}{2009}]{gur09}
{Gurzadyan} V.~G.,  {Allahverdyan} A.~E.,  {Ghahramanyan} T.,  {Kashin} A.~L.,
  {Khachatryan} H.~G.,  {Kocharyan} A.~A.,  {Kuloghlian} H.,  {Mirzoyan} S.,
  {Poghosian} E.,    {Yegorian} G.,  2009, \aap, 497, 343

\bibitem[\protect\citeauthoryear{{Gurzadyan} \& {Kocharyan}}{{Gurzadyan} \&
  {Kocharyan}}{2008}]{gur08}
{Gurzadyan} V.~G.,  {Kocharyan} A.~A.,  2008, \aap, 492, L33

\bibitem[\protect\citeauthoryear{{Hansen}, {Banday} \& {G{\'o}rski}}{{Hansen}
  et~al.}{2004}]{han04b}
{Hansen} F.~K.,  {Banday} A.~J.,    {G{\'o}rski} K.~M.,  2004, \mnras, 354, 641

\bibitem[\protect\citeauthoryear{{Hansen}, {Cabella}, {Marinucci} \&
  {Vittorio}}{{Hansen} et~al.}{2004}]{han04a}
{Hansen} F.~K.,  {Cabella} P.,  {Marinucci} D.,    {Vittorio} N.,  2004, \apjl,
  607, L67

\bibitem[\protect\citeauthoryear{{Hinshaw} et~al.,}{{Hinshaw}
  et~al.}{2009}]{hin09}
{Hinshaw} G.,  et~al., 2009, \apjs, 180, 225

\bibitem[\protect\citeauthoryear{{Hoftuft}, {Eriksen}, {Banday}, {G{\'o}rski},
  {Hansen} \& {Lilje}}{{Hoftuft} et~al.}{2009}]{hof09}
{Hoftuft} J.,  {Eriksen} H.~K.,  {Banday} A.~J.,  {G{\'o}rski} K.~M.,  {Hansen}
  F.~K.,    {Lilje} P.~B.,  2009, \apj, 699, 985

\bibitem[\protect\citeauthoryear{{Jarosik} et~al.,}{{Jarosik}
  et~al.}{2010}]{jar10}
{Jarosik} N.,  et~al., 2010, ArXiv e-prints

\bibitem[\protect\citeauthoryear{{Land} \& {Magueijo}}{{Land} \&
  {Magueijo}}{2005}]{lan05}
{Land} K.,  {Magueijo} J.,  2005, \mnras, 357, 994

\bibitem[\protect\citeauthoryear{{Larson} \& {Wandelt}}{{Larson} \&
  {Wandelt}}{2004}]{lar04}
{Larson} D.~L.,  {Wandelt} B.~D.,  2004, \apjl, 613, L85

\bibitem[\protect\citeauthoryear{{McEwen}, {Hobson}, {Lasenby} \&
  {Mortlock}}{{McEwen} et~al.}{2005}]{mce05}
{McEwen} J.~D.,  {Hobson} M.~P.,  {Lasenby} A.~N.,    {Mortlock} D.~J.,  2005,
  \mnras, 359, 1583

\bibitem[\protect\citeauthoryear{{Monteser{\'{\i}}n}, {Barreiro}, {Vielva},
  {Mart{\'{\i}}nez-Gonz{\'a}lez}, {Hobson} \& {Lasenby}}{{Monteser{\'{\i}}n}
  et~al.}{2008}]{mon08}
{Monteser{\'{\i}}n} C.,  {Barreiro} R.~B.,  {Vielva} P.,
  {Mart{\'{\i}}nez-Gonz{\'a}lez} E.,  {Hobson} M.~P.,    {Lasenby} A.~N.,
  2008, \mnras, 387, 209

\bibitem[\protect\citeauthoryear{{Mukherjee} \& {Wang}}{{Mukherjee} \&
  {Wang}}{2004}]{muk04}
{Mukherjee} P.,  {Wang} Y.,  2004, \apj, 613, 51

\bibitem[\protect\citeauthoryear{{Nolta} et~al.,}{{Nolta}
  et~al.}{2009}]{nol09}
{Nolta} M.~R.,  et~al., 2009, \apjs, 180, 296

\bibitem[\protect\citeauthoryear{{Pietrobon}, {Amblard}, {Balbi}, {Cabella},
  {Cooray} \& {Marinucci}}{{Pietrobon} et~al.}{2008}]{pie08}
{Pietrobon} D.,  {Amblard} A.,  {Balbi} A.,  {Cabella} P.,  {Cooray} A.,
  {Marinucci} D.,  2008, \prd, 78, 103504

\bibitem[\protect\citeauthoryear{{Pietrobon}, {Cabella}, {Balbi}, {Crittenden},
  {de Gasperis} \& {Vittorio}}{{Pietrobon} et~al.}{2010}]{pie10}
{Pietrobon} D.,  {Cabella} P.,  {Balbi} A.,  {Crittenden} R.,  {de Gasperis}
  G.,    {Vittorio} N.,  2010, \mnras, 402, L34

\bibitem[\protect\citeauthoryear{{R{\"a}th}, {Schuecker} \&
  {Banday}}{{R{\"a}th} et~al.}{2007}]{rae07}
{R{\"a}th} C.,  {Schuecker} P.,    {Banday} A.~J.,  2007, \mnras, 380, 466

\bibitem[\protect\citeauthoryear{{Rossmanith}, {R{\"a}th}, {Banday} \&
  {Morfill}}{{Rossmanith} et~al.}{2009}]{ros09}
{Rossmanith} G.,  {R{\"a}th} C.,  {Banday} A.~J.,    {Morfill} G.,  2009,
  \mnras, 399, 1921

\bibitem[\protect\citeauthoryear{{Schwarz}, {Starkman}, {Huterer} \&
  {Copi}}{{Schwarz} et~al.}{2004}]{sch04}
{Schwarz} D.~J.,  {Starkman} G.~D.,  {Huterer} D.,    {Copi} C.~J.,  2004,
  Physical Review Letters, 93, 221301

\bibitem[\protect\citeauthoryear{{Slosar} \& {Seljak}}{{Slosar} \&
  {Seljak}}{2004}]{slo04}
{Slosar} A.,  {Seljak} U.,  2004, \prd, 70, 083002

\bibitem[\protect\citeauthoryear{{Vielva}, {Mart{\'{\i}}nez-Gonz{\'a}lez},
  {Barreiro}, {Sanz} \& {Cay{\'o}n}}{{Vielva} et~al.}{2004}]{vie04}
{Vielva} P.,  {Mart{\'{\i}}nez-Gonz{\'a}lez} E.,  {Barreiro} R.~B.,  {Sanz}
  J.~L.,    {Cay{\'o}n} L.,  2004, \apj, 609, 22

\bibitem[\protect\citeauthoryear{{Vielva} \& {Sanz}}{{Vielva} \&
  {Sanz}}{2010}]{vie10}
{Vielva} P.,  {Sanz} J.~L.,  2010, \mnras, pp 458--+

\bibitem[\protect\citeauthoryear{{Vielva}, {Wiaux},
  {Mart{\'{\i}}nez-Gonz{\'a}lez} \& {Vandergheynst}}{{Vielva}
  et~al.}{2007}]{vie07}
{Vielva} P.,  {Wiaux} Y.,  {Mart{\'{\i}}nez-Gonz{\'a}lez} E.,
  {Vandergheynst} P.,  2007, \mnras, 381, 932

\bibitem[\protect\citeauthoryear{{Wiaux}, {Vielva}, {Barreiro},
  {Mart{\'{\i}}nez-Gonz{\'a}lez} \& {Vandergheynst}}{{Wiaux}
  et~al.}{2008}]{wia08}
{Wiaux} Y.,  {Vielva} P.,  {Barreiro} R.~B.,  {Mart{\'{\i}}nez-Gonz{\'a}lez}
  E.,    {Vandergheynst} P.,  2008, \mnras, 385, 939

\bibitem[\protect\citeauthoryear{{Wiaux}, {Vielva},
  {Mart{\'{\i}}nez-Gonz{\'a}lez} \& {Vandergheynst}}{{Wiaux}
  et~al.}{2006}]{wia06}
{Wiaux} Y.,  {Vielva} P.,  {Mart{\'{\i}}nez-Gonz{\'a}lez} E.,
  {Vandergheynst} P.,  2006, Physical Review Letters, 96, 151303

\end{thebibliography}

\end{document}